\documentclass[
  journal=ancac3,
  manuscript=article
]{achemso}

\usepackage{amsmath,amssymb,amsfonts}
\usepackage{bm}
\usepackage{color, soul}
\usepackage{graphicx}
\usepackage{float}
\usepackage{wrapfig}
\usepackage{verbatim}
\usepackage[english]{babel}
\usepackage[dvipsnames]{xcolor}
\usepackage[
  colorlinks=true,
  bookmarks=false,
  citecolor=blue,
  urlcolor=blue,
  linkcolor=blue,
]{hyperref}
\usepackage{natbib}
\usepackage{subfigure}
\usepackage{amstext, mathrsfs, textcomp}
\usepackage{multirow}
\usepackage{siunitx}
\usepackage{enumerate}
\usepackage{enumitem}
\usepackage{epstopdf}
\usepackage{tabularx}
\makeatletter

\usepackage[]{mdframed}
\mdfsetup{skipabove=20pt,skipbelow=10pt}

\newtheorem{theorem}{Theorem}

\newcommand{\qqoute}[1]{`#1'}

\usepackage{physics}
\usepackage{upgreek}

\newcommand{\iu}{\mathrm{i}\mkern1mu}

\newcommand{\bsans}[1]{\textbf{\textsf{#1}}}

\newcommand{\lcm}{\operatorname{lcm}}

\usepackage{gensymb}

\sethlcolor{yellow}

\setkeys{acs}{etalmode=truncate,maxauthors=0}
\graphicspath{{Figures/}}

\author{Ivan Sinev}
\affiliation{Institute of Bioengineering, \'Ecole Polytechnique F\'ed\'erale de Lausanne (EPFL), Lausanne 1015, Switzerland}
\altaffiliation{Contributed equally}
\author{Felix Ulrich Richter}
\affiliation{Institute of Bioengineering, \'Ecole Polytechnique F\'ed\'erale de Lausanne (EPFL), Lausanne 1015, Switzerland}
\altaffiliation{Contributed equally}
\author{Ivan Toftul}
\affiliation{Nonlinear Physics Center, Australian National University, Canberra ACT 2601, Australia}
\altaffiliation{Contributed equally}
\author{Nikita Glebov}
\affiliation{Institute of Bioengineering, \'Ecole Polytechnique F\'ed\'erale de Lausanne (EPFL), Lausanne 1015, Switzerland}
\author{Kirill Koshelev}
\affiliation{Nonlinear Physics Center, Australian National University, Canberra ACT 2601, Australia}
\author{Yongsop Hwang}
\affiliation{Laser Physics and Photonics Devices Lab, STEM, University of South Australia, Mawson Lakes, SA
5095, Australia}
\author{David G. Lancaster}
\affiliation{Laser Physics and Photonics Devices Lab, STEM, University of South Australia, Mawson Lakes, SA
5095, Australia}
\author{Yuri Kivshar}
\affiliation{Nonlinear Physics Center, Australian National University, Canberra ACT 2601, Australia}
\email{yuri.kivshar@anu.edu.au}
\author{Hatice Altug}
\affiliation{Institute of Bioengineering, \'Ecole Polytechnique F\'ed\'erale de Lausanne (EPFL), Lausanne 1015, Switzerland}
\email{hatice.altug@epfl.ch}

\title{Chirality encoding in resonant metasurfaces governed by lattice symmetries}

\begin{document}

\begin{abstract}

Chiral metasurfaces provide invaluable tools capable of controlling structured light required for biosensing, photochemistry, holography, and quantum photonics. Here we suggest and realize a universal strategy for controlling the chiral response of resonant metasurfaces \textit{via} the interplay of meta-atom geometry and lattice arrangements within all five possible planar Bravais symmetries. By introducing chiral gradient metasurfaces, we illustrate how our approach allows producing a predictable chiral response tunable by simple parameter variations. We highlight that symmetry-controlled chiral response provides an additional degree of freedom in optical signal processing, and showcase this with simultaneous mid-IR image encoding in two fundamental quantities, transmission and circular dichroism. Our proposed concept represents a universal toolkit for on-demand design and control of chiral metastructures that has potential for numerous applications in life sciences, quantum optics and more. 
\end{abstract}

\section{Introduction}

Chirality is a fundamental geometric property that describes a special type of asymmetry of an object, where it cannot be superimposed on its mirror image\cite{Kelvin1894Clarendon,Caloz2020IEEE,Caloz2020IEEE2,Barron2012Chirality}. Chirality plays a critical role in numerous fields, from chemistry and biology to solid state and quantum physics\cite{Smith2009ToxSciences,Kobayashi2011Nov,fasman2013circular,Aiello2022Apr}.

One of the most effective ways to manipulate and probe chiral systems is through the use of light, which can exhibit chiral properties, with circularly polarized (CP) plane waves being a straightforward example.
The selective interaction of light with chiral media manifests as difference in right and left CP light absorption (circular dichroism)\cite{Baase1979,fasman2013circular,Kobayashi2011Nov} or phase delay (circular birefringence)\cite{Nechayev2019Aug}. Employing chiral light enables the development of highly specific and efficient characterization systems for applications such as drug design, where the interaction between chiral molecules and biological systems can significantly affect efficacy and safety~\cite{Nguyen2006IJBS,Smith2009ToxSciences}. 
Beyond structural chirality, light can be used to probe the chiral properties of quantum mechanical systems, including orbital momentum~\cite{Bahramy2012NatCom}, spin, chiral wave functions, geometric phases, and topological phenomena~\cite{Lodahl2017NL}.
Additionally, chiral structures can be employed in counterfeit prevention or authentication methods, offering a high level of security through unique non-fungible optical signatures~\cite{Deng2020NanoLetters,Singh2024AppInterfAndMat}.

Enhancement and manipulation of chiral response, have been long-term goals in photonics, with applications spanning both far-field polarization control and augmenting the chiral light-matter interaction. 
A natural candidate for both these challenges are metasurfaces that already showed great prospects for light control in diverse applications, such as replacing bulky optical components\cite{Khorasaninejad2017S,Arbabi2023NP}, biosensing\cite{Zhang2021Nanophotonics,Tittl2018Science}, frequency conversion\cite{Krasnok2018MatToday,Jangid2023AdvMat}, optical computing\cite{Zhou2024Nanophotonics,hwang2016optical,Hu2024NatComm}, geometric phase manipulation\cite{Xie2021PRL,Guo2022PI}, and more. Likewise, metasurfaces proved to be an excellent tool for engineering chiral light\cite{Kim2021Jun}. In the far-field optics, chiral metasurfaces are the backbone of meta-holograms\cite{BalthasarMueller2017Mar} and devices for polarization conversion and polarimetry\cite{Zhu2013IEEE,Teng2019PhotRes,Shah2022ACSPhot,Ding2018ApplSci,Arbabi2018ACSPhot}.
At the same time, they can be used for generation of superchiral near-fields\cite{Tang2011Apr} for enhanced interaction with chiral molecules.\cite{Mohammadi2018Jul,Mohammadi2019Aug,Garcia-Santiago2022Jun} 

A number of strategies to achieve chiral response in metasurfaces have been suggested so far~\cite{Khaliq2023AdvOptMat,Deng2024NatNanoPhot}. Among these, one notable approach involves designing \textit{chiral meta-atoms} that break all the in-plane\footnote{Term \qqoute{in-plane} refers to the plane defined by a substrate.} mirror symmetries irrespective of their periodic arrangement. 
Breaking only the in-plane mirror symmetries results in differences exclusively between the transmission of right-to-left circular polarization and left-to-right circular polarization (cross-polarized transmissions), without affecting co-polarized transmission~\cite{Plum2009APL}.
However, this is not always sufficient, as, for example, in metasurfaces with three-fold rotational symmetry and higher, the cross-polarized transmission contrast is zero~\cite{Koshelev2024JO}.
The breaking of out-of-plane\footnote{Term \qqoute{out-of-plane} refers to normal direction with the respect to a substrate.} mirror symmetry is essential for achieving differences in co-polarized transmission between right- and left-circularly polarized light.
It is generally achieved either by the 3D geometry of the meta-atoms themselves~\cite{Arteaga2016Feb,Goerlitzer2020Jun,Gorkunov2020PRL,Tanaka2020ACSNano,Kosters2017ACSPhot,Esposito2015ACSPhot,Kaschke2015AOM} or by the presence of the substrate~\cite{Shi2022NatCom,Wang2023NanoPhot,Koshelev2024JO,Koshelev2023ACSPhot,Tonkaev2024NL}. 
Another strategy involves \textit{chiral lattices}: non-chiral meta-atoms arranged in a monoclinic lattice on a substrate\cite{Toftul2024PRL} or in a twisted bi-layered structures (moir\'e chiral photonic metasurfaces)~\cite{Lyu2023LPR,Wu2018Nanoscale,Han2023AM}. Furthermore, chiral response can be achieved if the metasurface is based on bianisotropic material with its anisotropy axis inclined with respect to the metasurface normal\cite{Asadchy2018Jun}. Lastly, the symmetry can also be broken externally by using non-normal light incidence~\cite{Cao2016OptMatExp,hwang2017optical,Nicolas2023ApplOptMat}, which induces the so-called extrinsic (or pseudo-) chirality~\cite{Nechayev2019Aug}. 
A combination of all strategies is also possible but would result in a system with a huge number of free parameters hard to design, optimize and often fabricate. At the same time, precise control of the chirality of the nanophotonics structures remains crucial both for generation of structured light and for reliable quantitative chiral detection\cite{Graf2019Feb,Gilroy2019Jun,Both2022Feb}.

In this paper, we suggest a universal strategy for controlling metasurface chiral response by leveraging the interplay of lattice and meta-atom symmetries to produce a \textit{predictable} chiral behavior tunable by simple parameters. While particular cases of such interaction were discussed before both in theory~\cite{Volkov2009Apr,Meng2022Oct,Movsesyan2022AOM,Avalos-Ovando2022ACSPhot} and experiment~\cite{Gryb2023NL}, here we develop \textit{a generalized framework} that addresses all possible symmetry combinations, serving as a universal toolkit for on-demand chiral design.

\begin{figure}[ht!]
  \centering
  \includegraphics[width=0.9\textwidth]{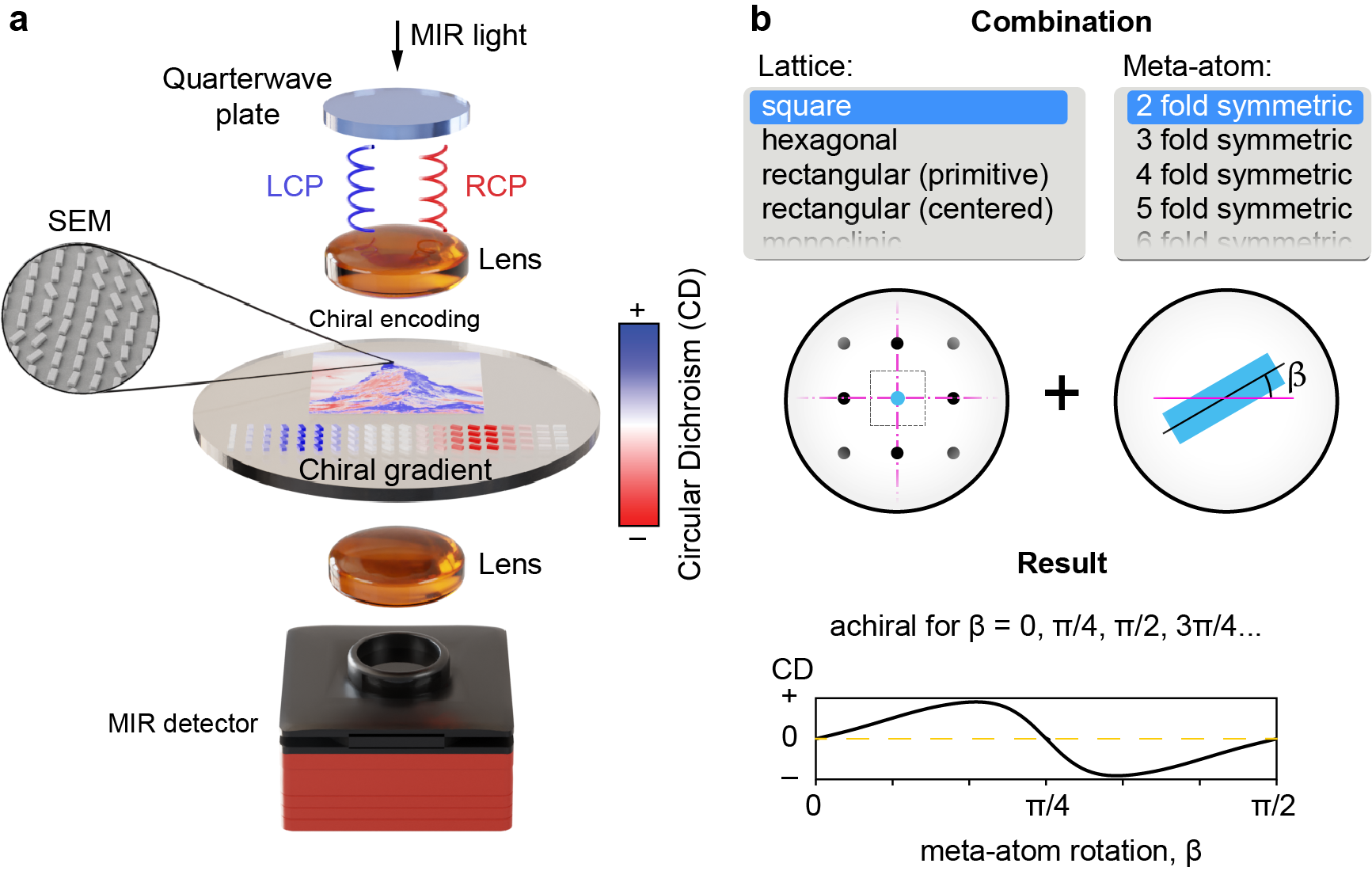}
  \caption{\textbf{Concept of chiral encoding using the interplay between lattice and meta-atoms symmetries}. \bsans{a} Artistic view of a chip hosting a chiral gradient metasurface and a metasurface encoding an image in circular dichroism signal in the mid-IR spectral range. The inset shows a tilted angle SEM image of a metasurface encoding a chiral image. \bsans{b} Schematic illustrating the interaction of the metasurface lattice symmetry with the resonator symmetry leading to a locked set of resonator rotation angles $\beta$ that render the structure non-chiral.}
  \label{fig:Concept} 
\end{figure}

We showcase this approach experimentally using \textit{chiral gradient metasurfaces}. In these structures, the rotation angle of the meta-atom within the unit cell is continuously varied along the chip, which allows to probe the full range of mutual orientations of lattice and resonator in the chiral signal measured with polarization-resolved mid-infrared spectroscopy (\autoref{fig:Concept}\bsans{a}). We observe robust zeros of the metasurface chirality for resonator rotation angles consistent with the chiral selection rules that we devise (\autoref{fig:Concept}\bsans{b}), as well as variable enhancement of the chirality at intermediate angles stemming from the metasurface resonant modes. 

Remarkably, our approach enables seamless integration of resonant behavior in the chiral response with the simplest resonator geometries, which in turn opens the opportunities for data encryption. We demonstrate its application for mid-IR image encoding in two fundamental quantities, unpolarized transmission and transmission circular dichroism (CD). Importantly, the symmetry-protected anchor points expand the encoding range to cover a full spectrum of negative to positive values of CD, as schematically shown in \autoref{fig:Concept}\bsans{a}.
The choice of different combinations of lattice and resonator symmetries determines the trade-off between the encoding ranges, allowing for precise control over the encoded data, while the design simplicity ensures its high spatial resolution. Furthermore, the amplitude-based encoding approach does not require additional polarization optics for data extraction.
This highlights the strength of our general chiral design toolkit and ensures its straightforward adaptability to a wide field of applications.

\section{Lattice vs. resonator symmetries for chirality control}
\label{sec:theory}

\begin{table*}[h]
\caption{\textbf{Chiral selection rules}. Angle $\beta$ is the relative angle between meta-atom symmetry and global arrangement, $s$ is an integer, and $\operatorname{lcm}(\,\cdot, \cdot)$ is the least common multiple of two integers. Gray parallelograms show unit cells of all five Bravais lattices in 2D. Table shows $\beta$ angles that correspond to achiral configurations, which we term ``anchor points''. Note that a substrate is essential to break out-of-plane mirror symmetry. }
\label{tab:selectionrules}
\includegraphics[width=1\linewidth]{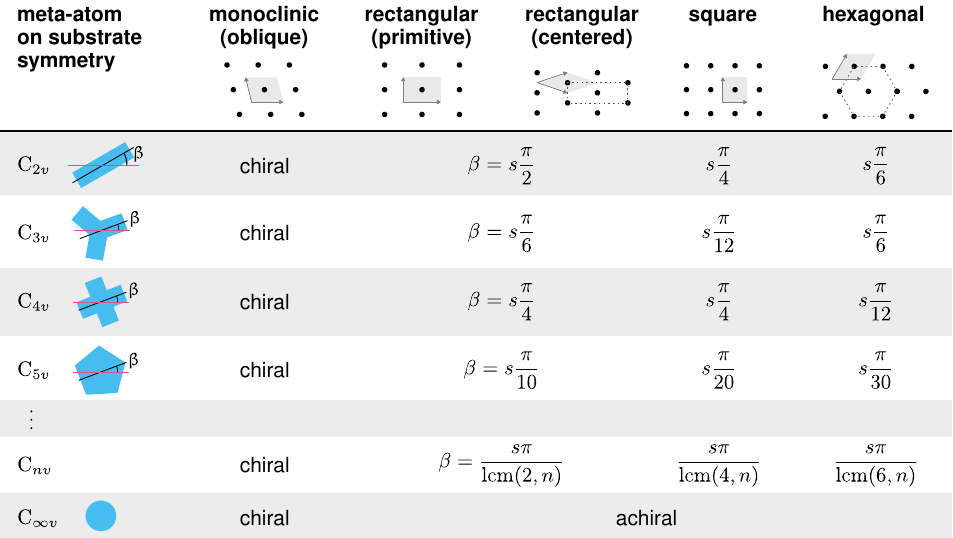}
\end{table*}

The core idea of our approach is the interplay between the 2D arrangement of the metasurface and the orientation of \textit{symmetric} meta-atoms with respect to this arrangement. By keeping the meta-atom shapes within simple non-chiral geometries, we can easily integrate resonant behavior into our system and achieve \textit{predictable chiral response} using achiral configurations as symmetry-protected anchor points.
In this section we devise the \textit{selection rules} which define the chiral response of such metasurfaces for all possible lattice types, \autoref{tab:selectionrules}.

Based on the geometrical definition of chirality~\cite{Caloz2020IEEE,Kelvin1894Clarendon}, a structure is chiral when it has no mirror symmetry planes. Let us consider the conditions for achieving \textit{achiral} response in our lattice --- that is, when at least one mirror symmetry plane appears. 
For the considered structures, we can discard the out-of-plane mirror symmetry, which is broken by default due to the presence of a substrate~\cite{Gorkunov2024AdvOptMat,Koshelev2024JO,Nechayev2019ACSPhot,Toftul2024PRL}. The chiral response is then solely dependent on the in-plane symmetries of our structure.
There are two symmetries we have to take into consideration: 
\textit{symmetry of the meta-atoms} and \textit{symmetry of the arrangement}. Without the loss of generality, we will consider meta-atoms which are $n$-fold symmetric and thus have $n$ in-plane mirror symmetry lines (such meta-atom on a substrate has $\mathrm{C}_{nv}$ point symmetry group). 
While a single $\mathrm{C}_{nv}$ object is achiral, this restriction can be lifted for a 2D periodic arrangement of $\mathrm{C}_{nv}$ objects, which opens the opportunity for precise chirality control that is unavailable for chiral meta-atoms.

In two dimensions, there are only five distinct variants of periodic lattices. These five 2D Bravais lattice types~\cite{Kittel2005} are  hexagonal (6), square (4), centered rectangular (2), primitive rectangular (2), and monoclinic (0), where the number in parentheses indicates number of in-plane mirror symmetry lines, $m$):. 
The only arrangement that is chiral for \textit{any} shape of meta-atoms on a substrate is monoclinic lattice, which was described theoretically and verified experimentally very recently~\cite{Toftul2024PRL}. However, it gives little room for an efficient chirality parametrization. 

Two sets of in-plane mirror symmetry lines ($m$ from lattice arrangement and $n$ from meta-atom) give us an easily parameterized system which is chiral for the particular relative orientations.
The relative angle of meta-atom rotation $\beta$ with respect to the arrangement lattice is the parameter of choice. By figuring out when exactly there is at least one match between symmetry lines from $m$-set and $n$-set, we derive the \textit{chiral selection rules}, \autoref{tab:selectionrules} (see Supplementary Information S1 for the detailed derivation). For specific angles $\beta$ the metasurface is achiral and hence must provide null for any chiral optical response.
The angular period of these \textit{symmetry-induced zeros} folds down to an elegant and compact formula
\begin{equation}
  \Delta \beta = \frac{\pi}{\lcm(m,n)},
  \label{eq:delta_beta}
\end{equation}
where $\lcm(m,n)$ is the least common multiple of two integers $m$ and $n$.
These symmetry induced zeros can serve as ``\textit{anchors points}'' for any chiral design in general, including metasurfaces with gradient design, which is the showcase of our choice in this work.
The later one is the showcase of our choice in this work. 

To characterize the chirality of our system, we use
\begin{equation}
  \text{CD}_{\text{\text{tot}}} = \frac{T_{\text{R}} - T_{\text{L}}}{T_{\text{R}} + T_{\text{L}}}
  \label{eq:CDtot}
\end{equation}
which can take values from $-1$ to $1$. 
Here, $T_{\text{R}}$ and $T_{\text{L}}$ denote the total transmission coefficients for the right and left circularly polarized input light, respectively.
The value defined by Eq.~\eqref{eq:CDtot} is related to both anisotropic dissymetry factor~\cite{Wakabayashi2014JPCA,Berova2007CSR} (with a factor of 2 difference) and to circular dichroism as defined for lossless metasurfaces~\cite{Koshelev2024JO,Tonkaev2024NL,Shalin2023}. 
Importantly, for 2D achiral metasurface configurations defined by \autoref{tab:selectionrules}, CD$_{\text{tot}}$ is zero and thus can be used as a measure of chirality in our design. In general, however, one should be careful when operating with total/co-polarized CD\footnote{Co-polarized circular dichroism is defined as $\mathrm{CD}_{\text{co}} = (T_{\text{RR}} - T_{\text{LL}}) / (T_{\text{RR}} + T_{\text{LL}})$ where $T_{\text{RR}}$ and $T_{\text{LL}}$ are the intensity transmission coefficients with the first and last indexes denoting the
output and input polarizations, correspondingly.} in cases when polarization conversion is critical~\cite{Koshelev2024JO,Toftul2024PRL,Shalin2023,Gorkunov2020PRL,Kondratov2016PRB}, as there is a possibility of $\text{CD}_{\text{tot}} \neq 0$ for 2D chiral (3D achiral) metasurfaces~\cite{Plum2009APL}. We discuss this difference in more details in Supplementary Section S2. For simplicity, hereafter we refer to CD$_{\text{tot}}$ as simply ``circular dichroism'' or CD.

\begin{figure}[ht!]
  \centering
  \includegraphics[width=\textwidth]{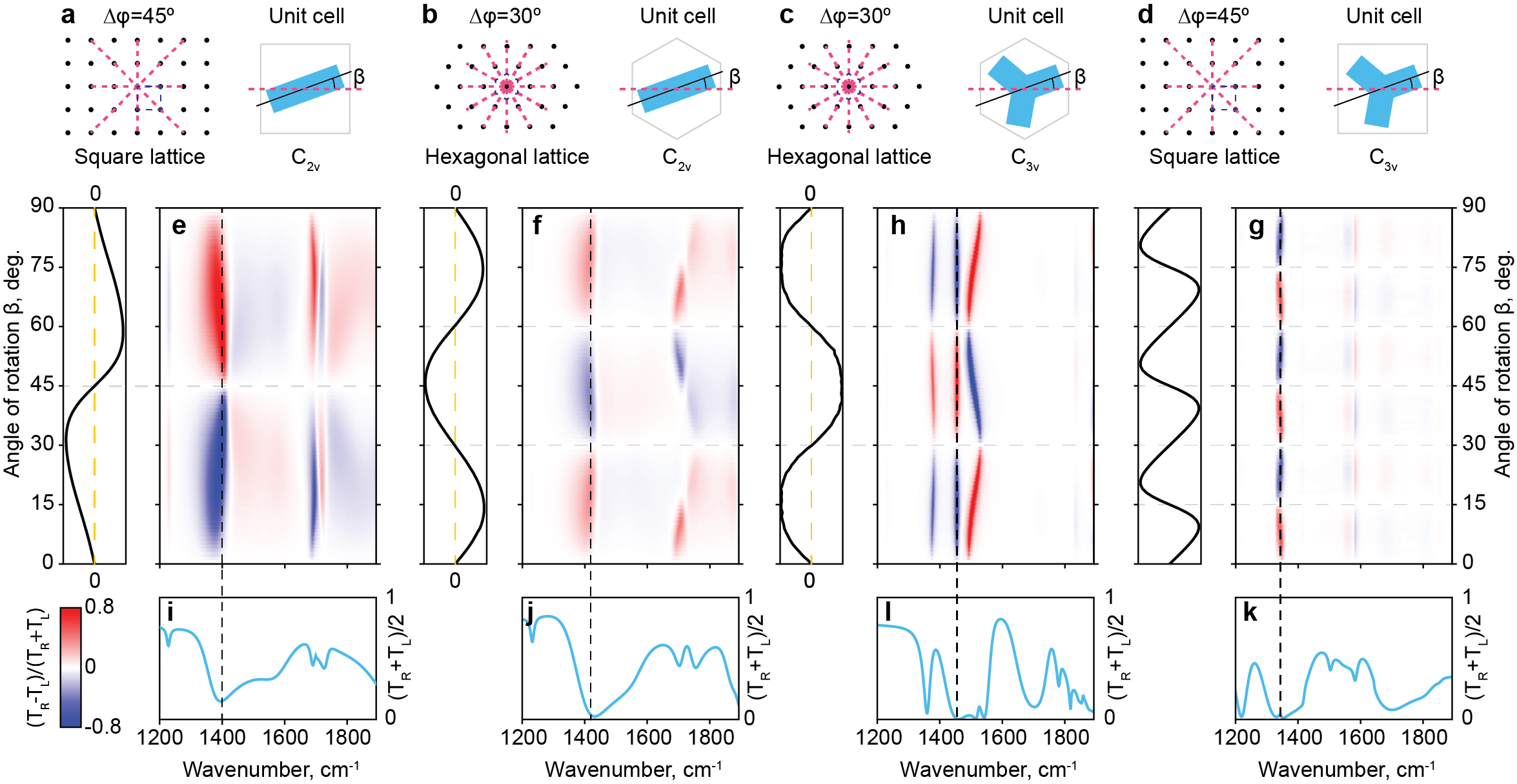}
  \caption{\textbf{Simulations of circular dichroism for different lattice and resonator symmetries.} \bsans{a}-\bsans{d} Schematic of the lattice symmetries and unit cell configuration for four type of metasurfaces: (\bsans{a},\bsans{b}) based on $\mathrm{C}_{2v}$ bar resonator and (\bsans{c},\bsans{d}) based on $\mathrm{C}_{3v}$ spinner resonator. \bsans{e}-\bsans{f} Corresponding calculated maps of circular dichroism (CD) depending on the excitation wavelength and rotation angle of the resonator. The maps demonstrate angle-equidistant zeroes governed by the interplay of the symmetries of the resonator and the lattice. The insets on the left of each map shows the section of CD taken at a wavelength marked with black dashed line in each of the panels (\bsans{e}-\bsans{h}). Bottom row: unpolarized light transmission spectra calculated for each type of metasurface for $\beta =0$. Horizontal dashed lines mark the $\beta$ corresponding to symmetry-protected zeroes of CD.}
  \label{fig:Simulations} 
\end{figure}

\section{Metasurfaces with gradient chirality}
\label{sec:chiral_grad}

The designs of our mid-IR chiral metasurfaces are based on Ge resonators on CaF$_2$ substrates, which was established as a flexible platform for light control in mid-IR\cite{Leitis2019SciAdv,Richter2024Jun}. To achieve strong resonances of the far-field chirality of the structure within the mid-IR range (1000--2000~cm$^{-1}$), we set the resonator height at 1070~nm. Furthermore, we focus on the two resonator symmetries: $\mathrm{C}_{2v}$ (``bar'') and $\mathrm{C}_{3v}$ (``spinner''), see the first two rows of \autoref{tab:selectionrules}. The lateral dimensions of the bar resonator are $3.542\times1.566$ microns, while the spinner is constructed from three bars of the same size rotated by 120 degrees with respect to each other and sharing a common origin. The period of both square and hexagonal lattices is 4.30~$\upmu$m (see schematics in Supplementary Figure~S2). These geometries exhibit well-defined and spectrally separated resonances with different Q-factors, which showcases the general applicability of our approach. A more detailed discussion on the structure design is presented in Supplementary Sections S6 and S7.

For the quantification of the chirality we use the circular dichroism (CD) introduced in~Eq.~\eqref{eq:CDtot}. \autoref{fig:Simulations}\bsans{a}-\bsans{d} shows the four combinations of the resonator (bar and spinner) and lattice symmetries (square and hexagonal) that we considered in numerical simulations. According to the selection rules (\autoref{tab:selectionrules}), these combinations should yield chiral zeroes for $\beta=s\pi/{4}$ ($\mathrm{C}_{2v}$/square, \autoref{fig:Simulations}a), $\beta=s\pi/{6}$ ($\mathrm{C}_{2v}$/hexagonal, \autoref{fig:Simulations}b), $\beta=s\pi/{6}$ ($\mathrm{C}_{3v}$/hexagonal, \autoref{fig:Simulations}c) and $\beta=s\pi/{12}$ ($\mathrm{C}_{3v}$/square, \autoref{fig:Simulations}\bsans{d}). To verify this, we calculated the maps of $\mathrm{CD}_{\text{tot}}$ for infinitely periodic metasurfaces for $\beta$ within $0$ to $\pi/2$ range excited from the top (air). These maps shown in \autoref{fig:Simulations}\bsans{e}-\bsans{h} demonstrate strict lack of chirality ($\mathrm{CD}=0$) for discrete sets of $\beta$ (horizontal dashed lines) fully consistent with the selection rules. Notably, these zeros hold irrespective of the resonant modes of the metasurface that manifest as peaks of CD at intermediate~$\beta$ values. The sections of the maps at the wavelengths close to the respective CD maxima shown to the left of each map indicate that CD is strictly anti-symmetric with respect to each of the $\beta$ ``anchors''. Finally, \autoref{fig:Simulations}\bsans{i}-\bsans{k} show the unpolarized light transmission spectra of the respective symmetry configurations calculated for $\beta=0$. The resonant modes of the metasurface are manifested as transmission dips. These spectral bands of low transmission also manifest the highest CD, as according to  definition in Eq.~\eqref{eq:CDtot} the maximal values of CD are achieved at the minima of either RCP or LCP transmission (shown in Supplementary Fig. S7). This also explains low values of CD outside the resonant bands of the metasurface.

\begin{figure}[ht!]
  \centering
  \includegraphics[width=\textwidth]{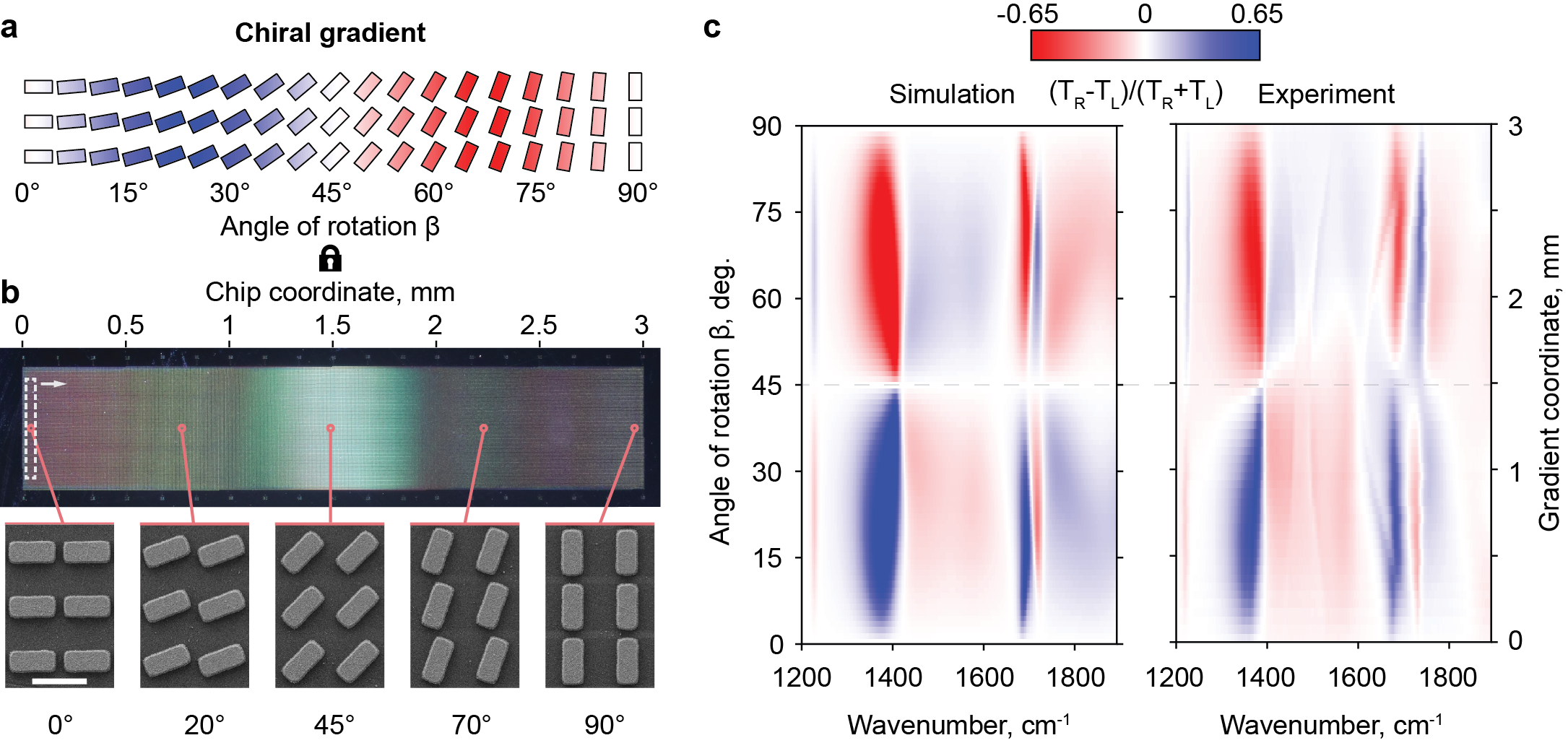}
  \caption{\textbf{Concept of chiral gradient metasurfaces.} \bsans{a} Schematic illustration of the chiral gradient metasurface with bar resonators arranged in a square lattice. The rotation angle $\beta$ of the resonator within the unit cell varies smoothly along the chip coordinate. The color of the bars encodes the chiral signal characteristic for the corresponding $\beta$. \bsans{b} Optical image of the fabricated chiral gradient metasurface based on Ge resonators on CaF$_2$ substrate. The insets show the SEM images, with six unit cells each, taken at different coordinates along the chiral gradient stripe. Gray dashed line indicates the signal collection area that is scanned along the gradient stripe during the measurement. \bsans{c} Comparison of simulated chiral signal map (left) and the experimentally measured dependence of chiral signal on the coordinate along the chip (right).}
  \label{fig:GradientExperiment}
\end{figure}

\begin{figure}[ht!]
  \centering
  \includegraphics[width=\textwidth]{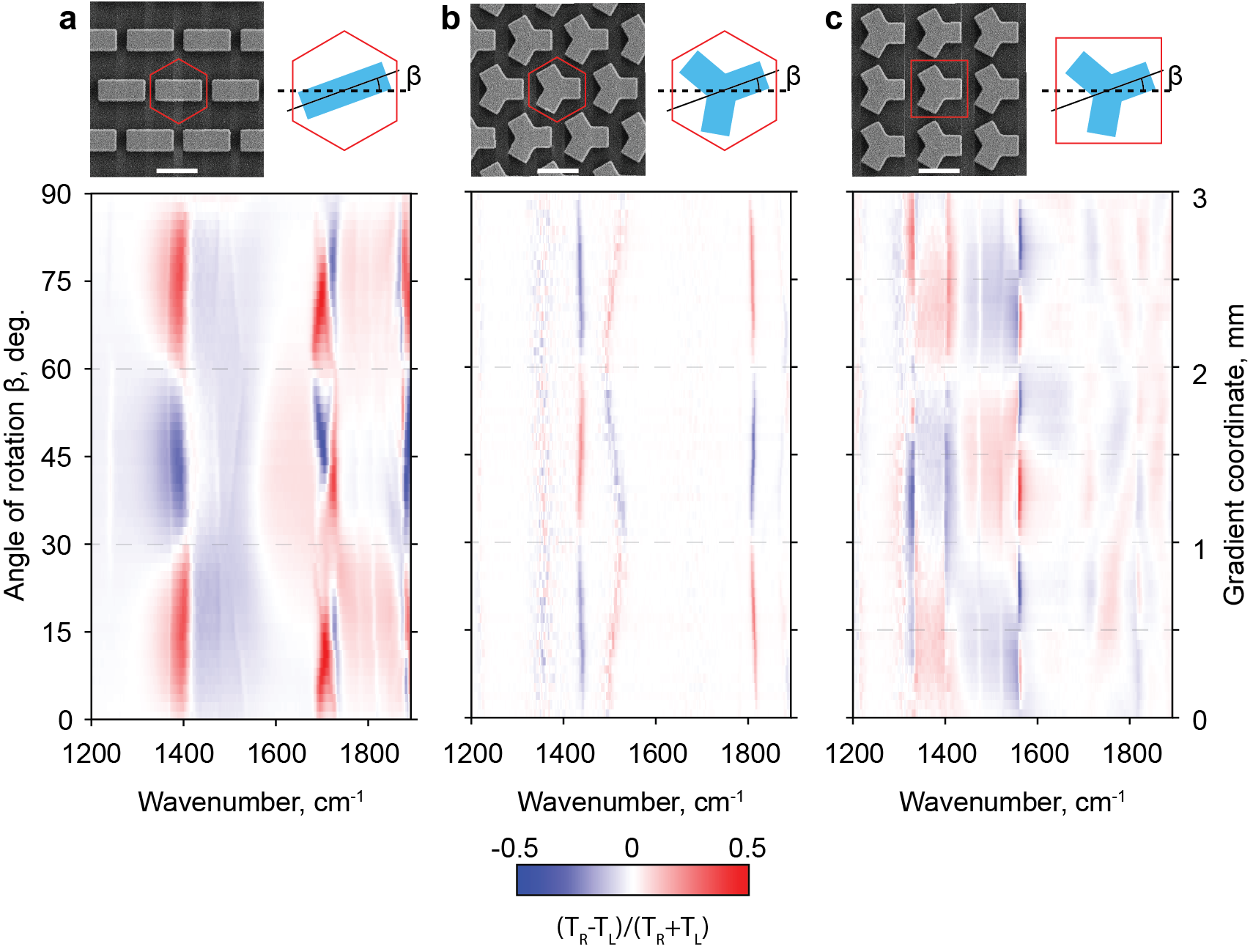}
  \caption{\textbf{Experimental results for custom combinations of resonator and lattice symmetries.} \bsans{a} $\mathrm{C}_{2v}$/hexagonal, \bsans{b} $\mathrm{C}_{3v}$/hexagonal, \bsans{c} $\mathrm{C}_{3v}$/rectangular. Top row: SEM images and unit cell schematics of the fabricated chiral gradient metasurfaces. Bottom row: experimentally measured spectral maps of chiral signal from the corresponding gradient metasurface samples.}
  \label{fig:customSymmetries}
\end{figure}

To showcase this general concept experimentally, we utilize \textit{chiral gradient} metasurfaces. Each of the fabricated metasurfaces addresses a certain combination of the resonator and lattice symmetries. However, instead of fabricating a discrete set of arrays for different angles $\beta$, we implemented a slow gradient of the rotation angle of the resonator along one of the chip coordinates. This yielded $600\times3000~\upmu\text{m}^2$ sized strips of gradient metasurfaces with $\beta$ varied within 0 to $\pi/2$ range along the long axis of the structure as schematically illustrated in \autoref{fig:GradientExperiment}\bsans{a}. Such scale results in $\Delta\beta\approx 0.1^{\circ}$ between the neighboring columns.

An optical image of the fabricated metasurface featuring bar ($\mathrm{C}_{2v}$) resonators in a square lattice is shown in \autoref{fig:GradientExperiment}\bsans{b}, with SEM insets highlighting its structural details at multiple points along the chiral gradient. Locking of the local value of $\beta$ with the chip coordinate of a gradient metasurface enables full characterization of the chiral properties of a given resonator-lattice symmetry pair through the measurements of the evolution of the LCP and RCP transmission spectra along the chip. We perform these measurements using a mid-IR microscope (Bruker Hyperion) in transmission configuration paired with Fourier transform infrared spectrometer (FTIR) as schematically illustrated in \autoref{fig:Concept}\bsans{a} (see Methods for further details). The experimental data of CD for $\mathrm{C}_{2v}$/square metasurface are in excellent agreement with our simulations (\autoref{fig:GradientExperiment}\bsans{c}) in all its features, including relative amplitude and spectral positions of the chirality peaks and, most importantly, robust zeroes of CD for $\beta=0, 45$ and $90$ degrees. We note that while the maximum chiral response is not the main focus of our study, CD for the proposed design reaches values larger than 0.8, which gives a dynamic range of 1.6 within the same structure. Additional narrow spectral features found in the experimental map in \autoref{fig:GradientExperiment}\bsans{c} are attributed to the modes excited at non-normal light incidence, which is present in the experiment due to finite numerical aperture of our mid-IR microscopy system (see Supplementary Section S4 for simulations).

\autoref{fig:customSymmetries} and Supplementary Figure S3 show the IR measurements of the gradient metasurfaces for the remaining lattice-resonator combinations showcased in numerical simulations (\autoref{fig:Simulations}\bsans{b}-\bsans{d}), in particular, $\mathrm{C}_{2v}$/hexagonal (\autoref{fig:customSymmetries}\bsans{a}), $\mathrm{C}_{3v}$/hexagonal (\autoref{fig:customSymmetries}\bsans{b}) and $\mathrm{C}_{3v}$/square (\autoref{fig:customSymmetries}\bsans{c}). The results for hexagonal lattice closely reproduce the numerical calculations and the expected chirality-canceling $\beta$ angles. Notably, the overall low transmission of both $\mathrm{C}_{3v}$ samples due to the overlap of multiple resonances within the 1400--1600~cm$^{-1}$ range leads to an increased noise level of the observed chiral features. 

A more complex picture of CD zeros is manifested for $\mathrm{C}_{3v}$/square structure (\autoref{fig:customSymmetries}\bsans{c}). The pronounced mode at $\approx 1570~\text{cm}^{-1}$ hosts 7 nodes with $\Delta \beta =15^{\circ}$ as expected from the simulation (\autoref{fig:Simulations}\bsans{g}). However, the two longer wavelength modes at $1320$ and $1400~\text{cm}^{-1}$ demonstrate nodes with $\Delta \beta =60^{\circ}$ instead (see also experimentally measured reflectivity maps in Supplementary S5). We attribute the suppression of the intermediate chirality zeroes to the contribution of oblique incidence in the experiment, leading to the manifestation of extrinsic chirality.

In contrast to the well-defined symmetry-protected anchor angles (Table~\ref{tab:selectionrules}), the behavior of CD at the intermediate angles --- particularly, the positions and the amplitudes of maximum values --- is less straightforward. The few approaches proposed so far still ultimately rely on numerical optimization procedures.\cite{Deng2024NatNanoPhot} For example, it is possible to estimate the transmission CD using the \textit{mode circular dichroism}, a property associated with the optical eigenmodes of the structure \cite{Gorkunov2020PRL,Toftul2024PRL,Gorkunov2024AdvOptMat}. Methods based on the eigenmode analysis have demonstrated their effectiveness in the vicinity of resonances, both for single high-$Q$ resonances \cite{Gorkunov2020PRL,Toftul2024PRL,Gromyko2024NC,Gromyko2024arXiv,Gorkunov2024AdvOptMat,Koshelev2023ACSPhot} and for broadband chiral responses achieved through multipolar superposition engineering \cite{Wang2023NanoPhot}.
Instead, our approach enforces a \textit{predictable} behavior of the chiral response controlled by a \textit{single} parameter, the meta-atom rotation angle. Importantly, the $\mathrm{C}_{nv}$ point symmetry of the meta-atom on a substrate guarantees a \textit{symmetric} range of CD values (apart from monoclinic arrangement). That is, for an optimized structure design exhibiting maximum chirality ($|\mathrm{CD}|=1$), the entire CD range $[-1, 1]$ is automatically available by variation of $\beta$, while an additional constraint of $\mathrm{C}_{nv}$ symmetry narrows down the parameter space for resonator design optimization. This distinguishes our platform from designs relying on inherently chiral meta-atoms, which require re-design to achieve the intermediate chirality values.

\begin{figure}[ht!]
  \centering
  \includegraphics[width=\textwidth]{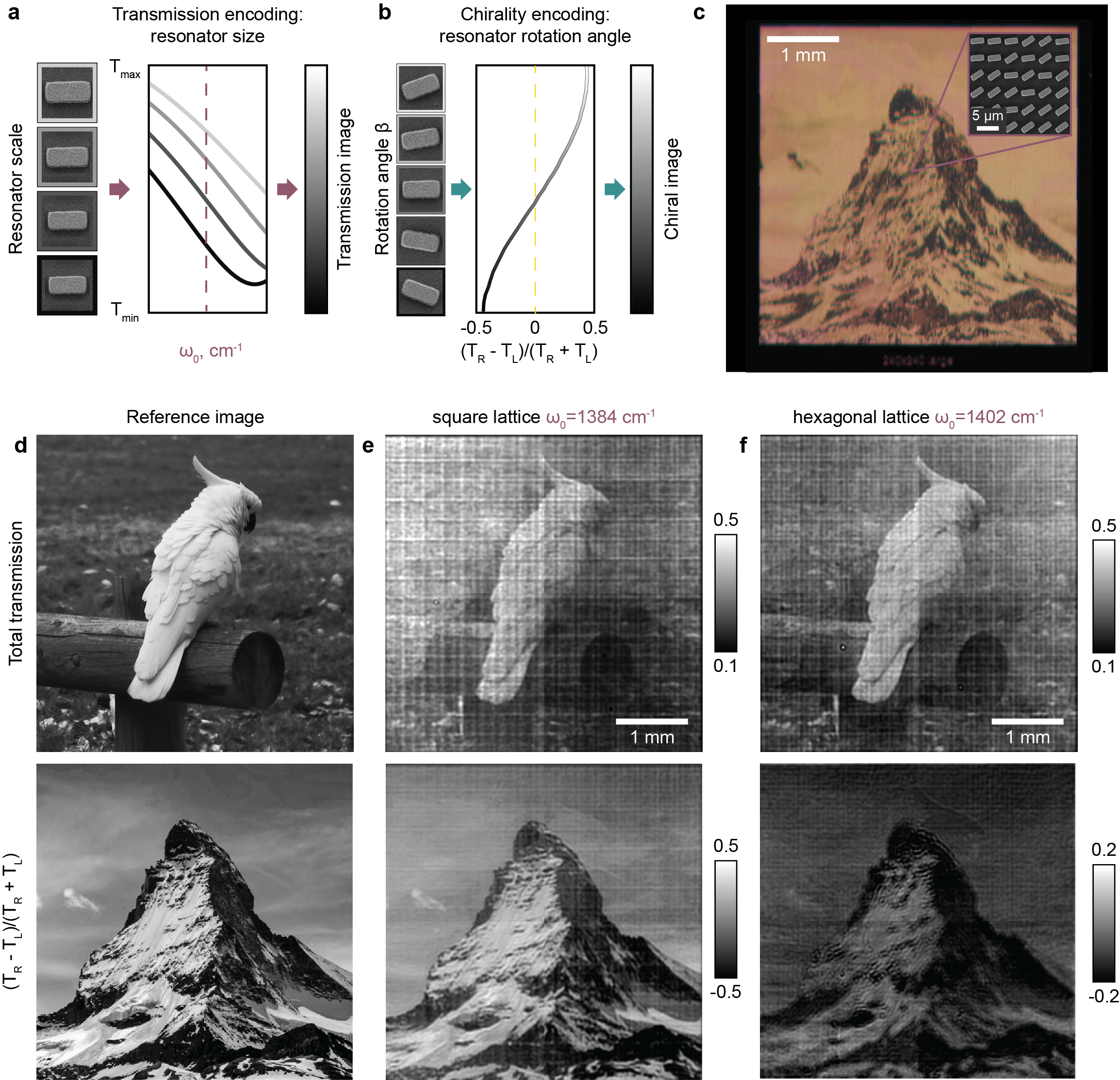}
  \caption{\textbf{Spatial variations of chirality for information encoding.} \bsans{a} Mechanism of encoding information in transmission image. Each scale of the resonator corresponds to a different level of transmission signal at a target wavelength $\omega_0$. \bsans{b} Mechanism of encoding information in chirality through changing the rotation angle $\beta$ of the resonator within a unit cell. \bsans{c} Optical image of the metasurface encoding two different images in transmission signal and chiral signal. The inset shows the SEM image of the metasurface area encoding $6\times6$ pixels of the original image. \bsans{d} Reference images. Top: a photo of an Australian cockatoo for encoding in transmission signal. Bottom: photo of Swiss Matterhorn mountain for chiral encoding. \bsans{e} Transmission (top row) and chiral signal (bottom row) images recorded from the encoding metasurface with square lattice design. \bsans{f} Respective images recorded from metasurface with hexagonal lattice design.}
  \label{fig:Encoding}
\end{figure}

One implication of the high dynamic range of CD offered by the symmetry interaction scheme is facilitated chiral information encoding. Contrary to the well-developed chiral holograms\cite{BalthasarMueller2017Mar}, where the metasurfaces are used to induce the near-field \textit{phase} profile that is then converted to a far-field image, our encoding principle is based on direct generation of signal \textit{amplitude} profile as illustrated in \autoref{fig:Encoding}\bsans{a},\bsans{b}. Specifically, we encode information in two fundamental quantities - total transmission and circular dichroism, which is possible with any resonator/lattice combination within our general model.  Each unit cell of a metasurface acts as a single pixel in an image encoded at a chosen frequency of light $\omega_0$, which is fundamentally different from other amplitude encoding approaches relying on interference effects of multiple elements\cite{Fan2020Dec}. The size of the resonator is used to encode data in total transmission signal, which is modulated due to the size-dependent spectral shift of the resonant modes. The resonator rotation angle is used to encode data in CD, as the detuning of $\beta$ from the symmetry nodes leads to the increase of the chirality. Reference grayscale images are then mapped on the available ranges $\Delta T$ and $\Delta \text{CD}$ so that each pixel (single unit cell) of a metasurface is assigned a particular combination of scale and $\beta$.  Notably, CD is not fully independent of the resonance shift, which leads to mutual mixing of the resonator scale and $\beta$ parameters in the encoding protocol and limits the total range of encoding depending on the lattice and resonator symmetries. This is illustrated in more detail in Supplementary Figure~S9.

We used two reference pictures --- of an Australian cockatoo and Swiss Matterhorn mountain (\autoref{fig:Encoding}\bsans{d}) --- to demonstrate chiral mid-IR image encoding in the total transmission and CD signals, respectively. As a showcase, we converted these two images into spatial maps of resonator scale and $\beta$ for metasurfaces based on $\mathrm{C}_{2v}$/square and $\mathrm{C}_{2v}$/hexagonal combinations, which show the largest tuning range of CD and transmission (see also Supplementary Fig.~S7), using the procedure described above. The image of the fabricated $\mathrm{C}_{2v}$/square sample in the visible light (\autoref{fig:Encoding}\bsans{c}) reveals the shape of the mountain, as the scattering of light mostly depends on the orientation of the bars, while the change of the size does not lead to noticeable modulation of reflectivity. In the mid-IR, where the resonant modes of the bars reside, the situation changes drastically. \autoref{fig:Encoding}\bsans{e},\bsans{f} show the images of two $\mathrm{C}_{2v}$ resonator metasurfaces with square and hexagonal lattices recorded at the excitation frequencies of 1384 and 1402~cm$^{-1}$, respectively, using a quantum cascade laser-based microscopy system (Daylight Solutions Spero, see Methods). The physical size of each metasurface was approximately $4 \times 4$~mm$^2$, which was tailored to fit into $2\times2$ fields of view of our microscope (this leads to barely noticeable stitching in the middle of the images). Comparison of the images for two lattice symmetries --- square (\autoref{fig:Encoding}\bsans{e}) and hexagonal (\autoref{fig:Encoding}\bsans{f}) --- highlights the advantages and drawbacks of using each option. Lower symmetry of square lattice leads to larger maximum amplitude of $\mathrm{CD}_{\text{tot}}$ for intermediate $\beta$ (\autoref{fig:Simulations}\bsans{e},\bsans{i}) that improves the chiral encoding, but at the same time limits the $\Delta T$ range, which influences the transmission encoding. Hexagonal lattice, while lacking in the maximum amplitude of chirality (\autoref{fig:Simulations}\bsans{f},\bsans{j}), provides a better balance of the encoding ranges, leading to good image quality in both signals. Additionally, the small angle range required to switch the sign of chirality for the hexagonal lattice bears the potential for dynamic encoding in active metasurfaces. This highlights the opportunities for more flexible control of chirality provided by custom lattice symmetries that can be tailored depending on particular application. Good quality of the images also indicates that local spatial variations of $\beta$ do not interfere with the lattice effect required to achieve the encoded values of CD.

\section{Conclusion}

In summary, we developed a universal approach to design 2D chiral metasurfaces based on symmetry considerations. We established general selection rules that define the mutual orientations of the lattice and the meta-atoms at which the chiral response of the metasurface vanishes. These zero-chirality ``anchors'' are protected by the interplay of the meta-atom and lattice symmetries and do not depend the presence of resonant optical modes. We demonstrate this powerful concept experimentally using mid-IR chiral gradient metasurfaces, where the mutual orientation of the resonator and the lattice was varied over the chip length, displaying the periodic vanishing of chirality according to the selection rules. While our results were obtained for dielectric metasurfaces, the approach is more broadly applicable for plasmonic metasurfaces as well~\cite{Wu2022AdvComposHybMater}. Furthermore, predictable behavior of circular dichroism enforced by the interplay of symmetries enables a unique opportunity for simultaneous information encoding in two fundamental quantities, transmission and circular dichroism. This feature is showcased through mid-IR image encoding, where a metasurface unveils distinct images in these two channels, paving the way for advanced anti-counterfeiting, camouflage and security applications. Our generalized chiral design toolkit opens new opportunities for control of structured light in the dynamic field of chiral photonics, with possible applications in polarimetry, optical computing, sensing, chemistry, and quantum photonics.

\section{Methods}
\subsection{Metasurface Fabrication}
Germanium (\text{Ge}) metasurfaces were fabricated on calcium difluoride (\(\text{CaF}_2\)) substrates. First, the material stack was prepared by subsequent sputtering of a 5~nm silicon oxide (\(\text{SiO}_2\)) adhesion layer and a 1070~nm Ge layer on top of a 1~mm thick CaF$_2$ chip. The inverted metasurface pattern was written using electron beam lithography (Raith EPBG500+) on a spin-coated single-layer PMMA (PMMA 495k A8) positive tone resist. After the development, the pattern was directly transferred to the Ge film using a fluorine-based dry plasma etching process (Alcatel AMS 200 SE).

\subsection{Infrared spectroscopy}
We obtained the infrared (IR) transmission spectra using a Bruker Vertex 80v FT-spectrometer with an attached IR Microscope (HYPERION 3000) equipped with a liquid nitrogen cooled MCT detector. The metasurfaces were excited from the air side using a ZnSe lens with the focal length of 25~mm mildly focusing circularly polarized IR light on the sample surface. Transmitted light was collected with another 25~mm lens equipped with an additional iris placed at its back focal plane. Closing the iris allowed for limiting the numerical aperture of the system down to approximately 0.06 and thus suppressing the unwanted signal from oblique excitation angles. We measured transmission spectra for both right-circularly polarized (RCP) and left-circularly polarized (LCP) generated using a broadband quarter-wave plate (model XCN13 from G\& H, 5--8~$\upmu$m). Signal collection area was limited to a thin stripe by a double-blade aperture placed in the conjugate image plane of the IR microscope. This collection area was then scanned along the gradient stripe (as in \autoref{fig:GradientExperiment}\bsans{b}) to obtain the wavenumber vs angle maps. The sample chamber was constantly purged with dry air to provide consistently low level of humidity. 

\subsection{Infrared images}
The images of metasurface encoding the data in transmission and chiral signals were collected with a quantum cascade laser based imaging microscope (Daylight SPERO-QT-Z) featuring a $480\times 480$ pixel microbolometer array. Separate LCP and RCP transmission images were collected by introducing a broadband quarter-wave plate ($2.5\text{--}7.0~\upmu \text{m}$ from B.Halle) into the collimated incident beam. The measurements were performed with a sample uniformly illuminated with a monochromatic (full width at half maximum $<0.1~\text{cm}^{-1}$) laser source at the encoding wavelength. LCP and RCP images were normalized to the corresponding transmission image of a bare CaF$_2$ substrate. The images shown in the manuscript were stitched together from four fields of view of the microscope each covering a $2\times2~\text{mm}^{2}$ area.

\subsection{Simulations}
We obtained transmission and $\mathrm{CD}$ using the Frequency Domain Solver in CST Studio Suite with circularly polarized light excitation. In the simulations, we set the dielectric permittivity of germanium to $\varepsilon_{\text{Ge}} = 17.45 + \iu 0.1668$ with imaginary part added to account for the effective losses due to grained structure of thermally evaporated amorphous Ge. The level of losses was chosen based on fitting the spectral shape and position of the resonances. The dielectric permittivity of calcium fluoride was $\varepsilon_{\text{CaF}_2} = 1.87$. By design, the resonator with $\mathrm{C}_{2v}$ symmetry had 1566 nm width, 3542 nm length and $1070$~nm height, arranged in a lattice with a period of 4.24 $\mu$m. The schematics and dimensions of $\mathrm{C}_{2v}$ and $\mathrm{C}_{3v}$ resonators are illustrated in Supplementary Figure~S2.

\section*{Acknowledgements}
We acknowledge the European Union's Horizon Europe Research and Innovation Programme under agreements 101046424 (TwistedNano) and 101070700 (MIRAQLS). This work was supported by the Swiss State Secretariat for Education, Research and Innovation (SERI) under contract numbers 22.00018 and 22.00081. Y.K. acknowledges support from the Australian Research Council (Grant No. DP210101292) and the International Technology Center Indo-Pacific (ITC IPAC) via Army Research Office (contract FA520923C0023). The authors acknowledge the use of nanofabrication facilities at the Center of MicroNano Technology of \'Ecole Polytechnique F\'ed\'erale de Lausanne. Y.H. and D.L. acknowledge support received from Electro Optic Systems Pty. Limited. 
Y.H. acknowledges support from the National Computational Infrastructure (NCI) by the Australian Government via Adapter Allocation Scheme for computing resources. 
I.T. thanks Kristina Frizyuk for the valuable discussions on the selection rules. I.S. and F.R. acknowledge Timothy Mann for help with early experiments.

\newpage

\begin{center}
\LARGE{\textbf{Chirality encoding in resonant metasurfaces governed by lattice symmetries}
\\{\color{gray} \small -- SUPPLEMENTARY INFORMATION --}
}
    
\end{center}

\section{Symmetric axis analysis}
\noindent The symmetric zeros occur when one of the axes of symmetry of the meta-atom 
coincides with one of the axes of symmetry of the lattice.
Assumming at least one pair of the symmetric axes are aligned, 
the rotational angular difference to make the next alignment 
can be derived.

\begin{figure}[htbp!]
  \includegraphics[width=0.80\textwidth]{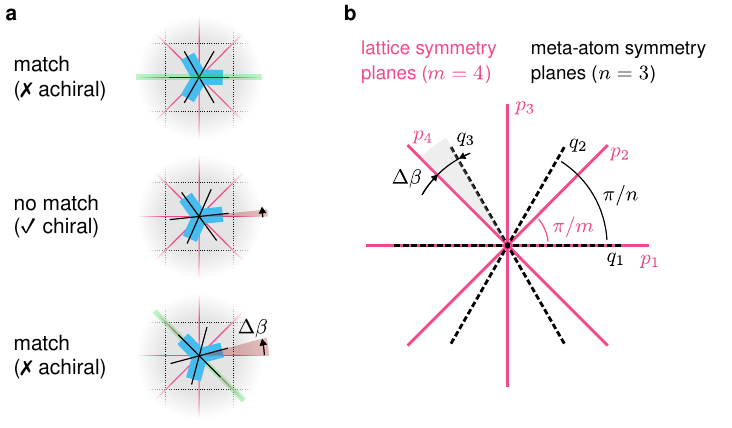}
  \caption{ 
  Illustration of meta-atom rotation with $\mathrm{C}_{3v}$ ($n=3$) in a square lattice with $m = 4$ in-plane mirror symmetry planes. Here $\Delta\beta$ is the achiral angular period. \bsans{a}  Visual representation of mirror symmetry plane matching during rotation. \bsans{b} Detailed illustration of all the relevant quantities used in the symmetry analyses.}
  \label{lcm}
\end{figure}

Consider the situation when the meta-atom is being rotated while the lattice remains fixed. If the lattice and the meta-atom have $m$ and $n$ axes of symmetry, respectively, the angles between the adjacent axes are $\pi/m$ and $\pi /n$, respectively.

Also let $p_i$ and $q_j$ represent the the symmetry axes, where the indices $i$ and $j$ increase counterclockwise, starting from the $x$-axis. Explicitly, the axes are labeled $p_1$ to $p_{m}$ for the lattice, and $q_1$ to $q_{n}$ for the meta-atom.
In \autoref{lcm} a particular case for a C$_{3v}$ meta-atom in a square lattice with $m=4$, $n=3$ as well as lattices $p_1$ to $p_4$ and $q_1$ to $q_3$ is illustrated.

Assume that $p_1$ and $q_1$ are initially aligned on the $x$-axis. We aim to determine the smallest rotational angle $\Delta \beta$ that brings the meta-atom into the next alignment of any pair of $p_i$ and $q_j$.
This is equivalent to finding the minimum angular difference of any two symmetry axes and can be expressed as $\left|\frac{\pi}{m} i- \frac{\pi}{n}j\right|$.

\begin{equation}
\Delta \beta = \left|\frac{\pi}{m} i- \frac{\pi}{n}j\right| 
           = \frac{|in - jm|\pi}{mn}.
\end{equation}

To find the minimum $\Delta \beta$ for a given pair of $(m, n)$, we need to minimize the absolute value in the numerator $|in - jm|$.
Assuming the greatest common divisor (GCD) of $m$ and $n$ is $a=\operatorname{gcd}(m,n)$, the two positive integers can be written as
\begin{equation}
  m=am^\prime,~\text{and}~ n=an^\prime,
\end{equation}
for the two coprime positive integers $m^\prime$ and $n^\prime$. Hence, the angle difference results in
\begin{equation}
\Delta \beta = \frac{|ian^\prime - jam^\prime|\pi}{a^2m^\prime n^\prime}
             = \frac{|in^\prime - jm^\prime|\pi}{am^\prime n^\prime}.
\end{equation}
which has its minimum when $|in^\prime - jm^\prime|=1$. Additionally, the denominator is the least common multiple (LCM) of $m$ and $n$: $\operatorname{lcm}(m,n)=am^\prime n^\prime$.

Now the question becomes whether it is possible to find a pair of $i$ and $j$ such that absolute value is 1.
According to B\'ezout's identity (below) we can always find a pair of $i$ and $j$ to make a difference of two coprime integers equal to 1.

\begin{mdframed}
\begin{theorem}[B\'ezout's identity]
\label{th:bezout}
Let $m$ and $n$ be integers with greatest common divisor $a$. Then there exist integers $i$ and $j$ such that $mi + nj = a$. 
\end{theorem}
\end{mdframed}

Since in our example $m^\prime$ and $n^\prime$ are coprime, their greatest common divisor is 1, which implies there exists a pair of $i$ and $j$ such that $|in^\prime - jm^\prime|=1$. Therefore, the minimum angle difference is
\begin{equation}
\Delta \beta_\text{min} = \frac{\pi}{am^\prime n^\prime}.
\end{equation}
This implies that the angular period for the symmetric zeros to appear is 
\begin{eqnarray}
\Delta \beta_\text{min} \nonumber 
  = \frac{\pi}{am^\prime n^\prime} \nonumber 
  = \frac{\pi}{\operatorname{lcm}(m,n)}. 
\end{eqnarray}

The number of zeros $N$ for the given interval of rotational angles can be found from the angular period of symmetric zeros. For example, in the rotational angle interval $[0, \pi /2]$
\begin{eqnarray}
  N = \frac{\pi/2}{\Delta \beta_\text{min}} \nonumber 
    = \frac{a m^\prime n^\prime}{2} \nonumber 
    = \frac{\operatorname{lcm}(m,n)}{2}.
\end{eqnarray}

\section{Circular dichroism and chirality}

Throughout the manuscript we are considering \textit{total} circular dichroism of the metasurfaces, which we define as $\text{CD}_{\text{tot}}=(T_{\text{R}}-T_{\text{L}})/(T_{\text{R}}+T_{\text{L}})$, with $T_R$ and $T_L$ denoting the \textit{total} transmission or right and left circularly polarized light, respectively. Operating with total transmission values greatly facilitates the measurements.
However, symmetry considerations only strictly define the zeros of the \textit{co-polarized} circular dichroism: $\text{CD}_{\text{co}}=(T_{\text{RR}}-T_{\text{LL}})/(T_{\text{RR}}+T_{\text{LL}})$.

Here, we calculate the cross-polarization components $T_\text{RL}$ and $T_{\text{LR}}$ of our metasurfaces and show that, for the considered combinations of lattice and resonator symmetries, their difference exhibits the same symmetry-defined zeros: $T_{\text{LR}}-T_{\text{RL}}=0$. Furthermore, for resonators with rotational symmetry of $n$-th order ($\mathrm{C}_{n}$) for $n\geq 3$, cross-polarized transmission and co-polarized reflection are prohibited. Hence, the afformentioned difference is zero even for intermediate angles (see Tab. 2 in Ref.~\cite{Koshelev2024JO} and Ch.~9 in Ref.~\cite{Shalin2023}).

\begin{figure}[H]
    \centering
    \includegraphics[width=\textwidth]{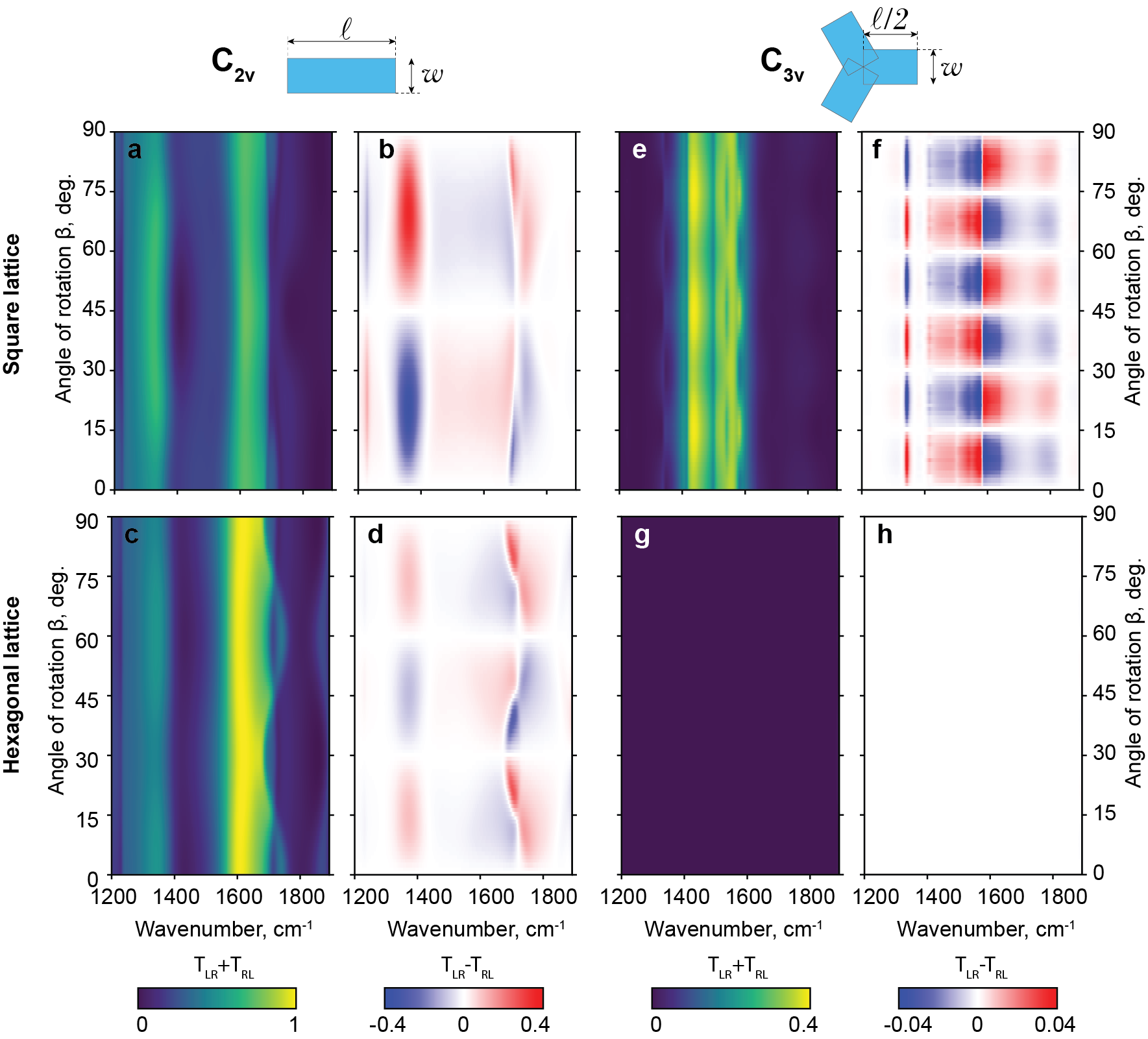}
    \caption{Top row: Schematic and geometric parameters of meta-atoms with C$_{2v}$ and C$_{3v}$ symmetry: $l = 3542$~nm and $w = 1566$~nm. Calculated maps of cross-polarized transmission spectra ($T_{LR} + T_{RL}$) and cross-polarized difference ($T_{LR} - T_{RL}$) for four types of metasurfaces considered in the main manuscript: \bsans{a}-\bsans{d} C$_{2v}$ resonator and \bsans{e}-\bsans{h} C$_{3v}$ resonator, each in the square and hexagonal lattice.}
    \label{fig:Cross_pol} 
\end{figure}

\section{Other lattice symmetries}
\begin{figure}[H]
    \centering
    \includegraphics[width=\textwidth]{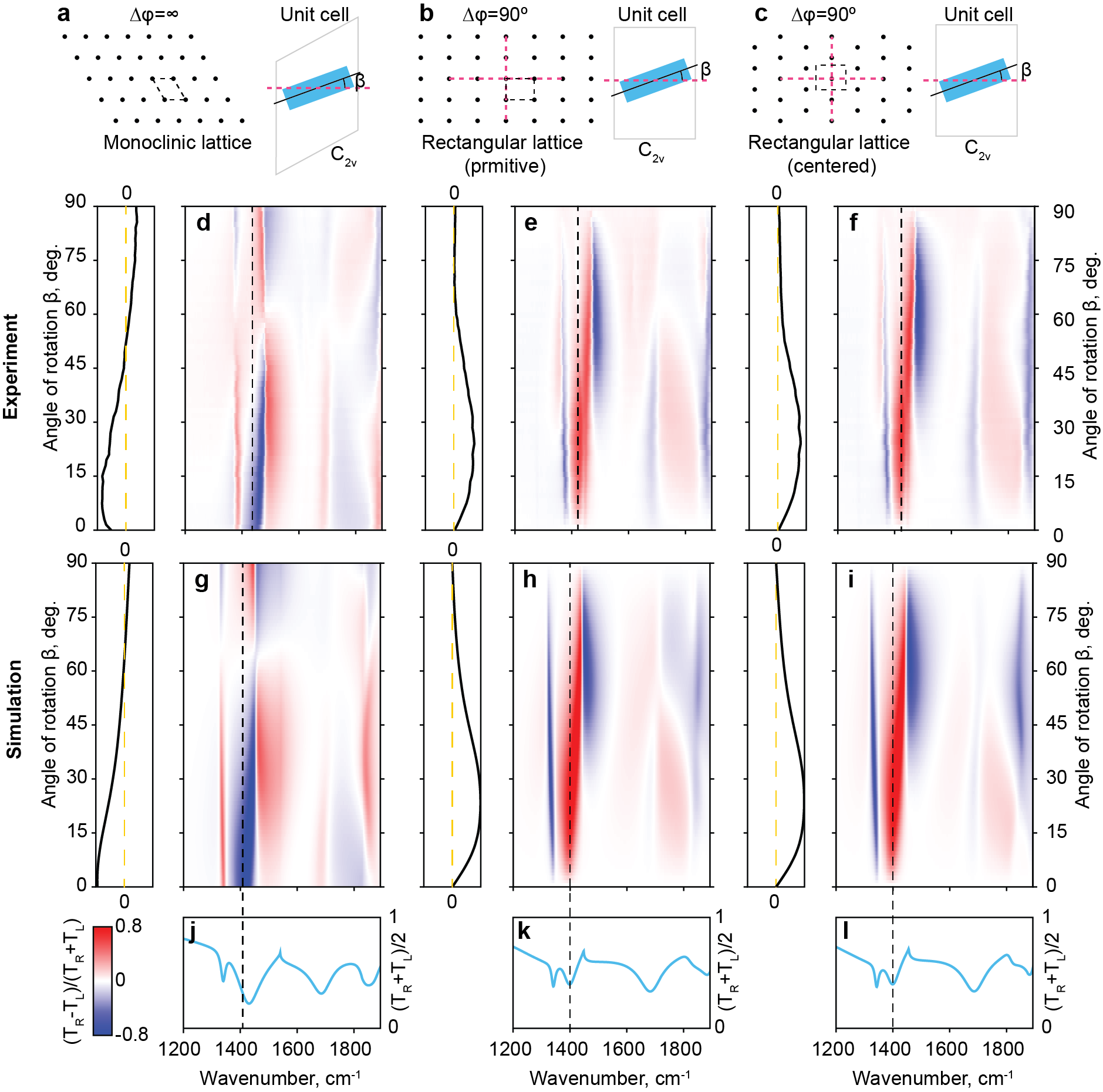}
    \caption{\bsans{a}-\bsans{c} Schematic of the lattice symmetries and unit cell configuration for three types of metasurfaces based on C$_{2v}$ bar resonator. \bsans{d}-\bsans{i} Corresponding \bsans{d}-\bsans{f} measured and \bsans{g}-\bsans{i} calculated maps of chiral signal depending on the excitation wavelength and rotation angle of the resonator. The insets on the left of each map shows the section of chiral signal taken at a wavelength marked with black dashed line in each of the panels \bsans{d}-\bsans{i}. \bsans{j}-\bsans{l} Unpolarized light transmission spectra calculated for each type of metasurface for $\beta$=0.}
    \label{fig:lowSymm} 
\end{figure}

Generally, a higher order of lattice symmetry leads to an increased number of chirality zeros within the tuning range of $\beta$, as detailed in Table~1 in the main manuscript. Here, we provide CD data for the remaining (low order) lattice symmetries -- monoclinic, rectangular primitive and rectangular centered -- not presented in the main text. \autoref{fig:lowSymm} compares experimental data measured from chiral gradient metasurfaces against simulations for a C$_{2v}$ resonators rotated in these three lattices. Minor spectral shifts of the modes are related to fabrication imperfections. Notably, the monoclinic lattice (\autoref{fig:lowSymm}\bsans{d},\bsans{e}) shows a decreased chirality near $\beta=70^{\circ}$ corresponding to the ``almost symmetric'' axis along the diagonal of the monoclinic unit cell. However, the simulation confirms that it is not a true chiral zero.

\section{Influence of the angle of incidence}

\begin{figure}[H]
    \centering
    \includegraphics[width=\textwidth]{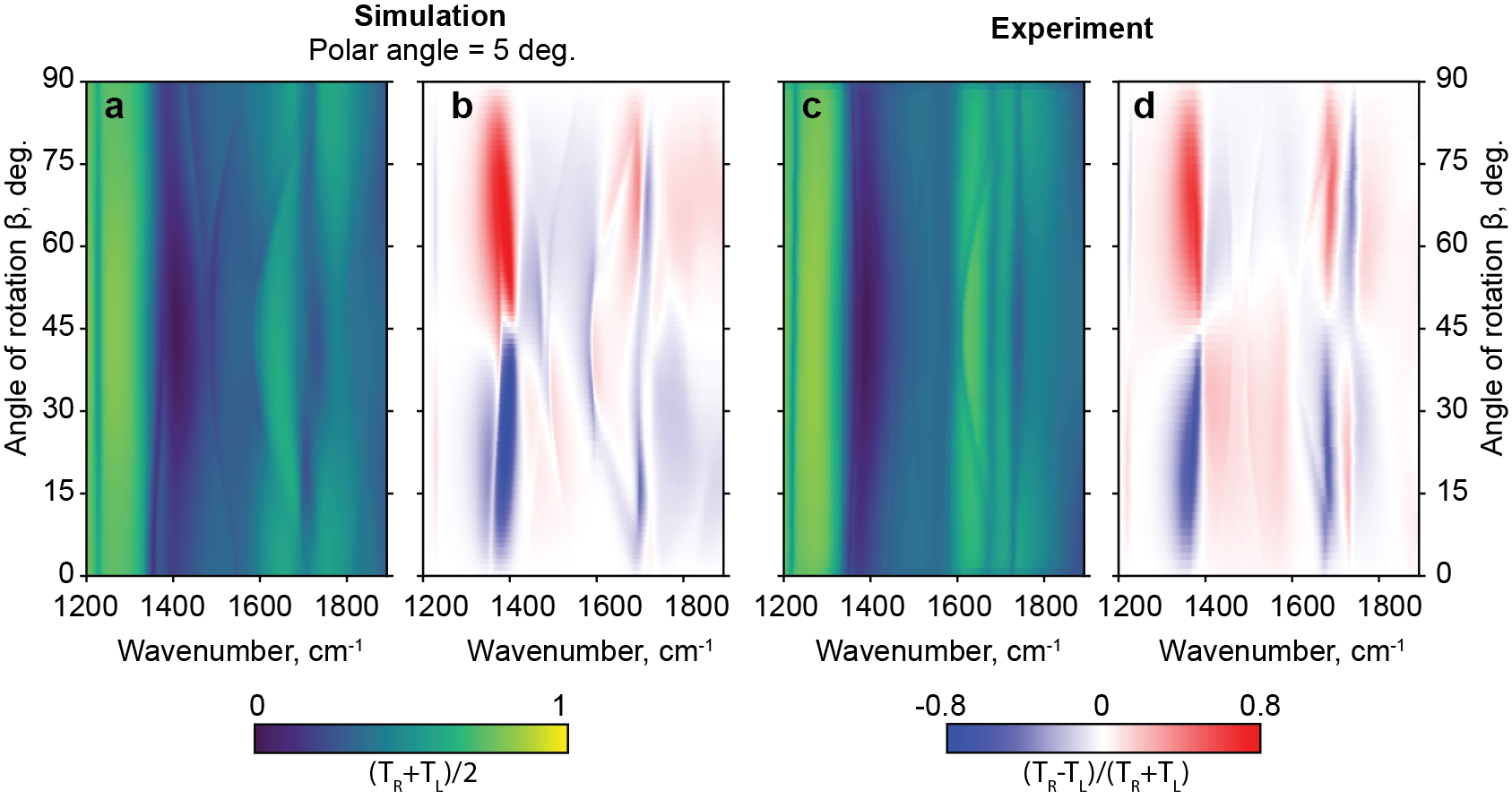}
    \caption{Maps of transmission spectra and chirality signal for C$_{2v}$/square lattice metasurface. \bsans{a},\bsans{b} Numerical calculation for 5 degrees angle of incidence. \bsans{c},\bsans{d} Experimentally measured data from a chiral gradient metasurface. The signal is collected within a numerical aperture of $\approx 0.06$. Maps in \bsans{a}-\bsans{d} contain similar sharp spectral features between 1300 and 1700~cm$^{-1}$ that correspond to optical modes excited at non-normal incidence.}
    \label{fig:oblique} 
\end{figure}

Experimentally measured maps of chiral gradient metasurfaces often contain narrow spectral features that are not present in the simulations at normal incidence. \autoref{fig:oblique} demonstrates that these features correspond to optical modes of the metasurfaces excited at non-normal incidence, as they are fully reproduced in the simulation data performed at a polar angle of 5 degrees (\autoref{fig:oblique}\bsans{a},\bsans{b}).

\section{Circular dichroism in reflection}

\begin{figure}[H]
    \centering
    \includegraphics[width=\textwidth]{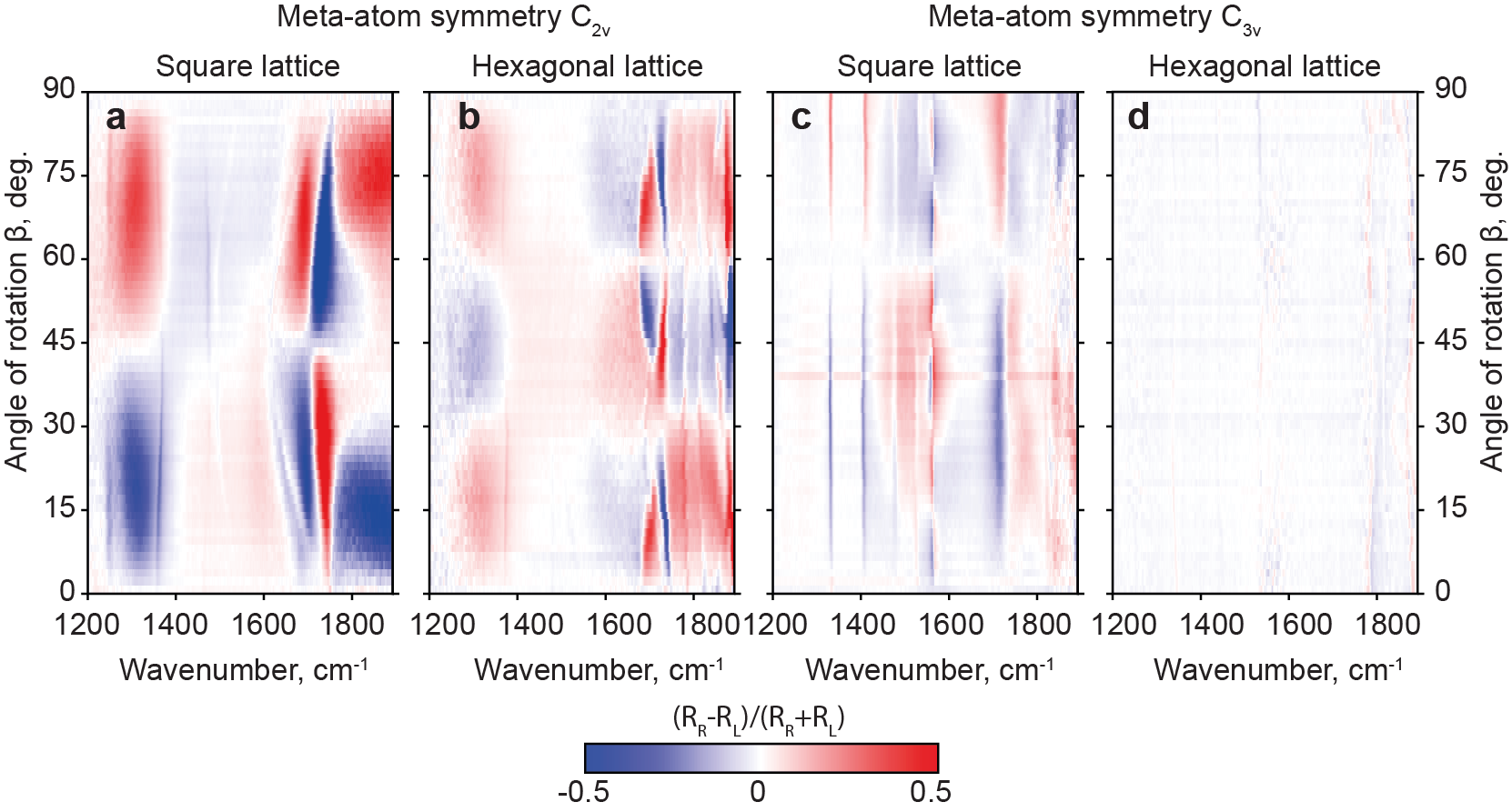}
    \caption{Maps of reflection chirality signal for four type of metasurfaces: based on \bsans{a},\bsans{b} C$_{2v}$ resonator and \bsans{c},\bsans{d} C$_{3v}$ resonator, each in the square and hexagonal lattice.}
    \label{fig:reflection} 
\end{figure}

Chiral features in transmission of $\mathrm{C}_{3v}$/square metasurface presented in the main manuscript suffer from increased noise level as they correspond to transmission minima defined by resonant modes of the metasurface. Here, we provide measurements of the reflection CD ($\text{CD}_{\text{R},\text{tot}} = (R_{\text{R}}-R_{\text{L}})/(R_{\text{R}}+R_{\text{L}})$) of this sample to capture the chiral features with better signal-to-noise ratio (\autoref{fig:reflection}\bsans{c}). For completeness, 
we also present reflection CD measurements for resonators with different symmetries and lattice structures (\autoref{fig:reflection}\bsans{a},\bsans{b},\bsans{d}).

\section{Electric field distributions and polarization-resolved transmission}

\begin{figure}[H]
    \centering
    \includegraphics[width=\textwidth]{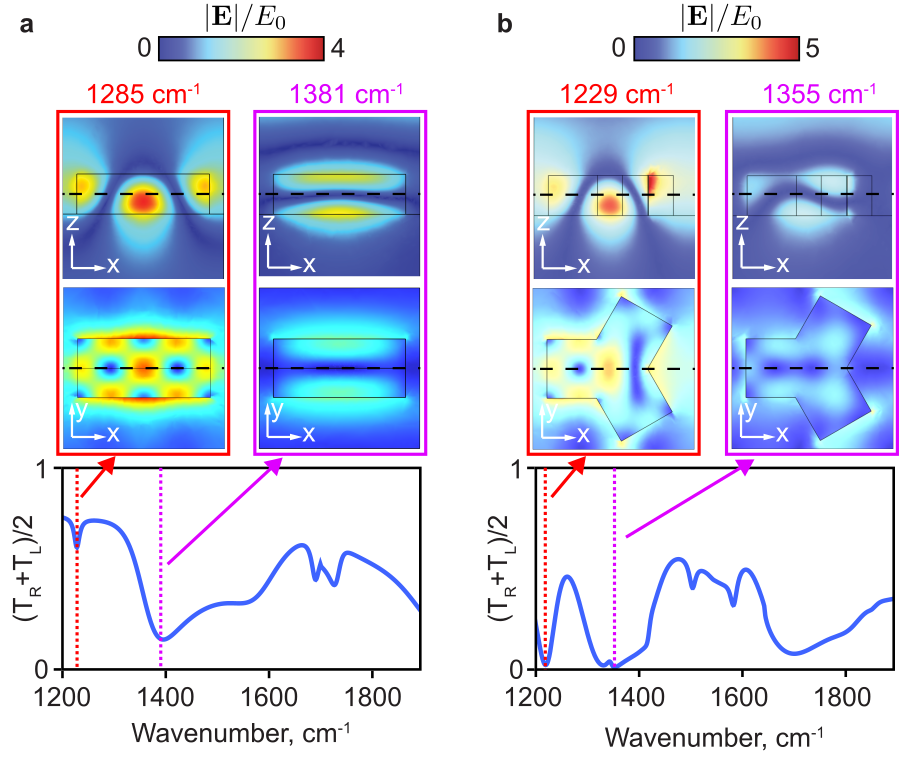}
    \caption{Calculated electric field distributions $|\mathbf{E}|$ for frequencies corresponding to resonant modes of $\mathrm{C}_{2v}$ (\bsans{a}) and $\mathrm{C}_{3v}$ (\bsans{b}) resonators in square lattice excited by a plane wave of amplitude $E_0$,  linearly polarized along the $y$ axis. Dashed lines show the positions of orthogonal sections ($xy$-section in $xz$-map and vice versa). Resonant frequencies at which the maps are plotted are shown with vertical lines in unpolarized transmission spectra shown in the bottom row. }
    \label{fig:field_profiles} 
\end{figure}

To gain more insight into the nature of the modes excited in the C$_{2v}$ and C$_{3v}$ resonators used as basic building blocks for our chiral structures, we plot the electric field distribution for linearly polarized plane wave excitation at frequencies corresponding to the lower energy resonances in transmission spectra. The electric field maps are shown in Fig.~\ref{fig:field_profiles}.

\begin{figure}[H]
    \centering
    \includegraphics[width=\textwidth]{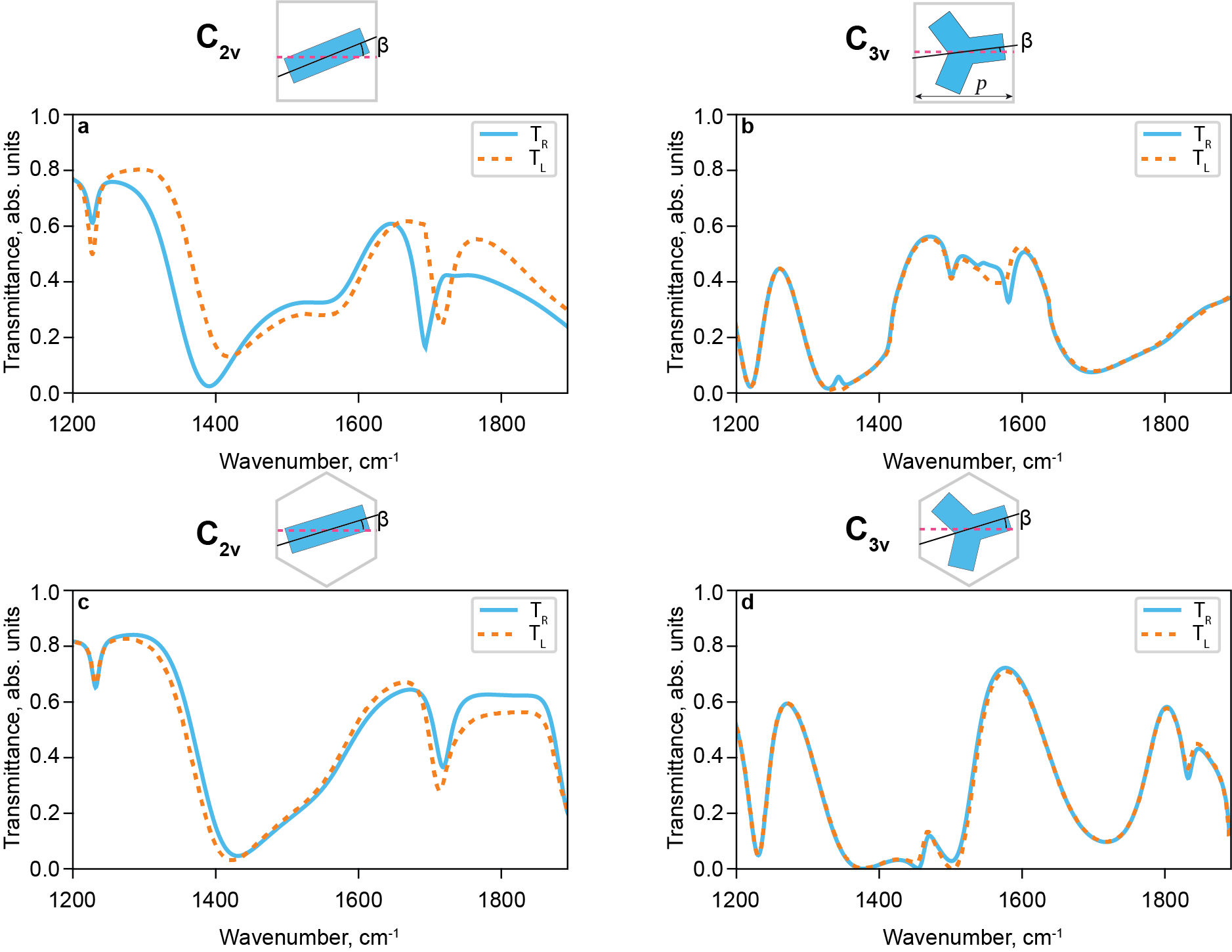}
    \caption{Calculated transmission spectra for the right- and left-circularly polarized light ($T_R$ and $T_L$) for metasurfaces based on (\bsans{a},\bsans{c}) C$_{2v}$ resonator and (\bsans{b},\bsans{d}) C$_{3v}$ resonator, each in the square and hexagonal lattices. The spectra are shown for $\beta$ values corresponding to the highest values of CD.}
    \label{fig:trans_rl} 
\end{figure}

Furthermore, to illustrate the amplitude of the chiral contrast leading to the manifestation of CD in the structures from the main text, in Fig.~\ref{fig:trans_rl} we separately plot the transmission spectra for left and right circularly polarized light. C$_{2v}$/square structure demonstrates the largest transmission contrast, with T$_R$-T$_L$ reaching 0.4. 

\section{Dependence on the lattice period}
\begin{figure}[H]
    \centering
    \includegraphics[width=\textwidth]{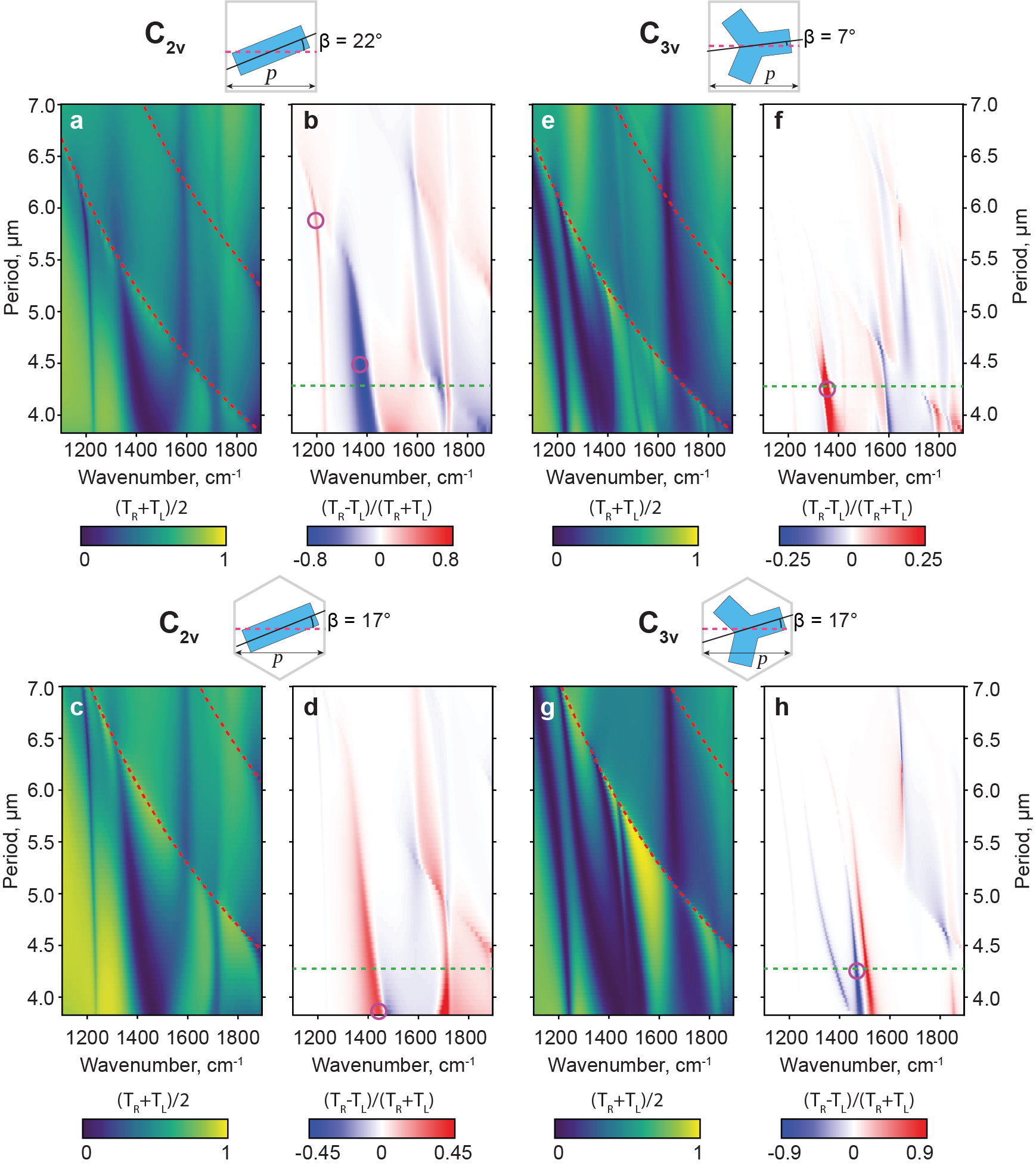}
    \caption{Calculated maps of transmission spectra $(T_{R} + T_{L})/2$ and CD signal $(T_{R} - T_{L})/(T_{R} + T_{L})$ depending on the lattice period $p$. The rotation angles $\beta$ correspond to the orientation with maximal chirality for the structures presented in the main manuscript with period of 4.3~µm shown with green dashed line in panels \bsans{b},\bsans{d},\bsans{f},\bsans{h}. \bsans{a}-\bsans{d} C$_{2v}$ resonator in the square ($\beta = 22\degree$) and hexagonal ($\beta = 17\degree$) lattice, \bsans{e}-\bsans{h} C$_{3v}$ resonator in the square ($\beta = 7\degree$) and hexagonal ($\beta = 17\degree$) lattice. The circles indicate maximum CD for a particular mode at the selected $\beta$. Diffraction cutoffs for substrate and air are shown with red dashed lines in panels \bsans{a},\bsans{c},\bsans{e},\bsans{g}.}
    \label{fig:trans_diff_period} 
\end{figure}

The designs of the resonators used in the main text (C$_{2v}$ bar and C$_{3v}$ spinner) are chosen to provide well-defined and spectrally separated resonances, simplifying the analysis and interpretation of the transmission and CD spectra. On the other hand, the period of the structure effectively controls the interaction strength of the meta-atoms in the lattice and, hence, can be an important tuning parameter to maximize the CD for intermediate values of resonator rotation angle $\beta$. \autoref{fig:trans_diff_period} shows the simulated spectra of unpolarized light transmission and CD for periods between 3.9 and 7~µm for four combinations of resonator and lattice symmetries considered in the main text. The maps reveal that chirality peaks corresponding to different modes manifest maximal CD at different periods.

Throughout the paper, we used a fixed period of 4.3 µm for three reasons: First, all four designs share a strong CD active mode at around 1400cm$^-1$ that is well manifested at the chosen period. Second, fixing the period removes the need to disentangle effects for multiple parameters and, thus, ensures a reliable comparison between different designs. Third, using a period of 4.3~µm leaves a minimal distance of 400~nm between the elements in the resonator arrays and allows for repeatable high quality nanofabrication compatible with large-area fabrication methods. Additionally, the difference between the CD at the optimal conditions and CD for the chosen period does not exceed a few percent.

Further optimization of the resonator design for particular application (for example, high-Q or broadband chirality) requires numerical optimization approaches\cite{Gorkunov2020PRL,Wang2023NanoPhot}.

\section{Chirality encoding}

\begin{figure}[h!]
    \centering
    \includegraphics[width=\textwidth]{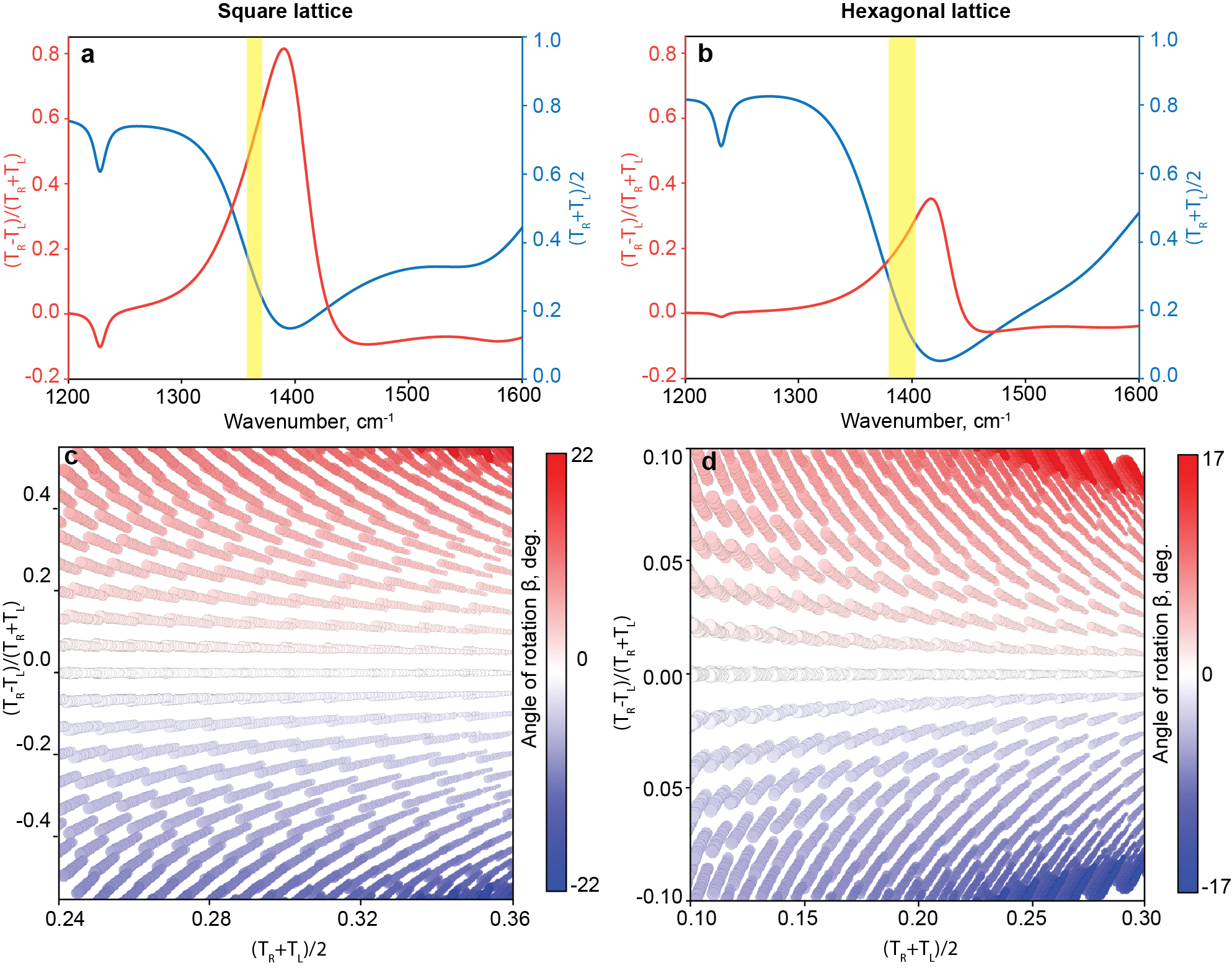}
    \caption{Calculated unpolarized light transmission spectra for $\beta =0$ and chiral signal for $\beta$ corresponding to the maximum chirality for metasurfaces based on  $\mathrm{C}_{2v}$ bar resonator in the \bsans{a} square and \bsans{b} hexagonal lattices. Yellow bars show the region of coding in transmission spectra. \bsans{c}-\bsans{d} Calculated chiral signal and transmission at the resonance excitation wavelength for varying resonator rotation angles and sizes.}
    \label{fig:Encode_data} 
\end{figure}
We calculated the transmission and chiral signals at the resonant excitation frequencies for various resonator rotation angles and sizes, to encode data in square and hexagonal lattices, respectively. The encoding wavelength was selected based on the high contrast between transmission and chiral signals (\autoref{fig:Encode_data}\bsans{a},\bsans{b}). Adjusting the rotation angle alters the chiral signal from minimum to maximum with minimal impact on transmission. The rotation angle varies from zero to the angle corresponding to the maximum chiral signal  ($\beta=22$ degrees for the square lattice and $\beta =17$ for the hexagonal lattice). The resonator size ranges from $L = 1500$ to $1680$~nm and from $w = 3450$ to $3530$~nm, without altering their ratio. Changes in resonator size shift the spectra and affect the transmission value at the resonant frequency. The transmission and chiral signal values fall within the required encoding region (\autoref{fig:Encode_data}\bsans{c},\bsans{d}). The intensity of each pixel in the two reference images is encoded by a point on the maps corresponding to specific resonator sizes and rotation angle. Finally, the pixels, each consisting of one resonator, are arranged within the lattice to form a metasurface.

\bibliography{Bibliography-General}

\providecommand{\latin}[1]{#1}
\makeatletter
\providecommand{\doi}
  {\begingroup\let\do\@makeother\dospecials
  \catcode`\{=1 \catcode`\}=2 \doi@aux}
\providecommand{\doi@aux}[1]{\endgroup\texttt{#1}}
\makeatother
\providecommand*\mcitethebibliography{\thebibliography}
\csname @ifundefined\endcsname{endmcitethebibliography}  {\let\endmcitethebibliography\endthebibliography}{}
\begin{mcitethebibliography}{82}
\providecommand*\natexlab[1]{#1}
\providecommand*\mciteSetBstSublistMode[1]{}
\providecommand*\mciteSetBstMaxWidthForm[2]{}
\providecommand*\mciteBstWouldAddEndPuncttrue
  {\def\EndOfBibitem{\unskip.}}
\providecommand*\mciteBstWouldAddEndPunctfalse
  {\let\EndOfBibitem\relax}
\providecommand*\mciteSetBstMidEndSepPunct[3]{}
\providecommand*\mciteSetBstSublistLabelBeginEnd[3]{}
\providecommand*\EndOfBibitem{}
\mciteSetBstSublistMode{f}
\mciteSetBstMaxWidthForm{subitem}{(\alph{mcitesubitemcount})}
\mciteSetBstSublistLabelBeginEnd
  {\mcitemaxwidthsubitemform\space}
  {\relax}
  {\relax}

\bibitem[Kelvin(1894)]{Kelvin1894Clarendon}
Kelvin,~W.~T. \emph{The molecular tactics of a crystal}; Clarendon Press, 1894; p~27\relax
\mciteBstWouldAddEndPuncttrue
\mciteSetBstMidEndSepPunct{\mcitedefaultmidpunct}
{\mcitedefaultendpunct}{\mcitedefaultseppunct}\relax
\EndOfBibitem
\bibitem[Caloz and Sihvola(2020)Caloz, and Sihvola]{Caloz2020IEEE}
Caloz,~C.; Sihvola,~A. {Electromagnetic Chirality, Part 1: The Microscopic Perspective [Electromagnetic Perspectives]}. \emph{IEEE Antennas Propag. Mag.} \textbf{2020}, \emph{62}, 58--71\relax
\mciteBstWouldAddEndPuncttrue
\mciteSetBstMidEndSepPunct{\mcitedefaultmidpunct}
{\mcitedefaultendpunct}{\mcitedefaultseppunct}\relax
\EndOfBibitem
\bibitem[Caloz and Sihvola(2020)Caloz, and Sihvola]{Caloz2020IEEE2}
Caloz,~C.; Sihvola,~A. {Electromagnetic Chirality, Part 2: The Macroscopic Perspective [Electromagnetic Perspectives]}. \emph{IEEE Antennas Propag. Mag.} \textbf{2020}, \emph{62}, 82--98\relax
\mciteBstWouldAddEndPuncttrue
\mciteSetBstMidEndSepPunct{\mcitedefaultmidpunct}
{\mcitedefaultendpunct}{\mcitedefaultseppunct}\relax
\EndOfBibitem
\bibitem[Barron(2012)]{Barron2012Chirality}
Barron,~L.~D. {From Cosmic Chirality to Protein Structure: Lord Kelvin's Legacy}. \emph{Chirality} \textbf{2012}, \emph{24}, 879--893\relax
\mciteBstWouldAddEndPuncttrue
\mciteSetBstMidEndSepPunct{\mcitedefaultmidpunct}
{\mcitedefaultendpunct}{\mcitedefaultseppunct}\relax
\EndOfBibitem
\bibitem[Smith(2009)]{Smith2009ToxSciences}
Smith,~S.~W. Chiral Toxicology: It’s the Same Thing…Only Different. \emph{Toxicological Sciences} \textbf{2009}, \emph{110}, 4–30\relax
\mciteBstWouldAddEndPuncttrue
\mciteSetBstMidEndSepPunct{\mcitedefaultmidpunct}
{\mcitedefaultendpunct}{\mcitedefaultseppunct}\relax
\EndOfBibitem
\bibitem[Kobayashi and Muranaka(2011)Kobayashi, and Muranaka]{Kobayashi2011Nov}
Kobayashi,~N.; Muranaka,~A. \emph{{Circular Dichroism and Magnetic Circular Dichroism Spectroscopy for Organic Chemists}}; Royal Society of Chemistry, 2011\relax
\mciteBstWouldAddEndPuncttrue
\mciteSetBstMidEndSepPunct{\mcitedefaultmidpunct}
{\mcitedefaultendpunct}{\mcitedefaultseppunct}\relax
\EndOfBibitem
\bibitem[Fasman(2013)]{fasman2013circular}
Fasman,~G.~D. \emph{Circular dichroism and the conformational analysis of biomolecules}; Springer Science \& Business Media, 2013\relax
\mciteBstWouldAddEndPuncttrue
\mciteSetBstMidEndSepPunct{\mcitedefaultmidpunct}
{\mcitedefaultendpunct}{\mcitedefaultseppunct}\relax
\EndOfBibitem
\bibitem[Aiello \latin{et~al.}(2022)Aiello, Abendroth, Abbas, Afanasev, Agarwal, Banerjee, Beratan, Belling, Berche, Botana, Caram, Celardo, Cuniberti, Garcia-Etxarri, Dianat, Diez-Perez, Guo, Gutierrez, Herrmann, Hihath, Kale, Kurian, Lai, Liu, Lopez, Medina, Mujica, Naaman, Noormandipour, Palma, Paltiel, Petuskey, Ribeiro-Silva, Saenz, Santos, Solyanik-Gorgone, Sorger, Stemer, Ugalde, Valdes-Curiel, Varela, Waldeck, Wasielewski, Weiss, Zacharias, and Wang]{Aiello2022Apr}
Aiello,~C.~D.; Abendroth,~J.~M.; Abbas,~M.; Afanasev,~A.; Agarwal,~S.; Banerjee,~A.~S.; Beratan,~D.~N.; Belling,~J.~N.; Berche,~B.; Botana,~A.; Caram,~J.~R.; Celardo,~G.~L.; Cuniberti,~G.; Garcia-Etxarri,~A.; Dianat,~A.; Diez-Perez,~I.; Guo,~Y.; Gutierrez,~R.; Herrmann,~C.; Hihath,~J.; Kale,~S.; Kurian,~P.; Lai,~Y.-C.; Liu,~T.; Lopez,~A.; Medina,~E.; Mujica,~V.; Naaman,~R.; Noormandipour,~M.; Palma,~J.~L.; Paltiel,~Y.; Petuskey,~W.; Ribeiro-Silva,~J.~C.; Saenz,~J.~J.; Santos,~E. J.~G.; Solyanik-Gorgone,~M.; Sorger,~V.~J.; Stemer,~D.~M.; Ugalde,~J.~M.; Valdes-Curiel,~A.; Varela,~S.; Waldeck,~D.~H.; Wasielewski,~M.~R.; Weiss,~P.~S.; Zacharias,~H.; Wang,~Q.~H. {A Chirality-Based Quantum Leap}. \emph{ACS Nano} \textbf{2022}, \emph{16}, 4989--5035\relax
\mciteBstWouldAddEndPuncttrue
\mciteSetBstMidEndSepPunct{\mcitedefaultmidpunct}
{\mcitedefaultendpunct}{\mcitedefaultseppunct}\relax
\EndOfBibitem
\bibitem[Baase and Johnson(1979)Baase, and Johnson]{Baase1979}
Baase,~W.~A.; Johnson,~W.~C. {Circular dichroism and DNA secondary structure}. \emph{Nucleic Acids Res.} \textbf{1979}, \emph{6}, 797--814\relax
\mciteBstWouldAddEndPuncttrue
\mciteSetBstMidEndSepPunct{\mcitedefaultmidpunct}
{\mcitedefaultendpunct}{\mcitedefaultseppunct}\relax
\EndOfBibitem
\bibitem[Nechayev \latin{et~al.}(2019)Nechayev, Barczyk, Mick, and Banzer]{Nechayev2019Aug}
Nechayev,~S.; Barczyk,~R.; Mick,~U.; Banzer,~P. {Substrate-Induced Chirality in an Individual Nanostructure}. \emph{ACS Photonics} \textbf{2019}, \emph{6}, 1876--1881\relax
\mciteBstWouldAddEndPuncttrue
\mciteSetBstMidEndSepPunct{\mcitedefaultmidpunct}
{\mcitedefaultendpunct}{\mcitedefaultseppunct}\relax
\EndOfBibitem
\bibitem[Nguyen \latin{et~al.}(2006)Nguyen, He, and Pham-Huy]{Nguyen2006IJBS}
Nguyen,~L.~A.; He,~H.; Pham-Huy,~C. {Chiral Drugs: An Overview}. \emph{International Journal of Biomedical Science : IJBS} \textbf{2006}, \emph{2}, 85\relax
\mciteBstWouldAddEndPuncttrue
\mciteSetBstMidEndSepPunct{\mcitedefaultmidpunct}
{\mcitedefaultendpunct}{\mcitedefaultseppunct}\relax
\EndOfBibitem
\bibitem[Bahramy \latin{et~al.}(2012)Bahramy, King, de~la Torre, Chang, Shi, Patthey, Balakrishnan, Hofmann, Arita, Nagaosa, and Baumberger]{Bahramy2012NatCom}
Bahramy,~M.; King,~P.; de~la Torre,~A.; Chang,~J.; Shi,~M.; Patthey,~L.; Balakrishnan,~G.; Hofmann,~P.; Arita,~R.; Nagaosa,~N.; Baumberger,~F. Emergent quantum confinement at topological insulator surfaces. \emph{Nature Communications} \textbf{2012}, \emph{3}\relax
\mciteBstWouldAddEndPuncttrue
\mciteSetBstMidEndSepPunct{\mcitedefaultmidpunct}
{\mcitedefaultendpunct}{\mcitedefaultseppunct}\relax
\EndOfBibitem
\bibitem[Lodahl \latin{et~al.}(2017)Lodahl, Mahmoodian, Stobbe, Rauschenbeutel, Schneeweiss, Volz, Pichler, and Zoller]{Lodahl2017NL}
Lodahl,~P.; Mahmoodian,~S.; Stobbe,~S.; Rauschenbeutel,~A.; Schneeweiss,~P.; Volz,~J.; Pichler,~H.; Zoller,~P. Chiral quantum optics. \emph{Nature} \textbf{2017}, \emph{541}, 473–480\relax
\mciteBstWouldAddEndPuncttrue
\mciteSetBstMidEndSepPunct{\mcitedefaultmidpunct}
{\mcitedefaultendpunct}{\mcitedefaultseppunct}\relax
\EndOfBibitem
\bibitem[Deng \latin{et~al.}(2020)Deng, Deng, Guan, Tao, Li, Li, Li, Yu, and Zheng]{Deng2020NanoLetters}
Deng,~J.; Deng,~L.; Guan,~Z.; Tao,~J.; Li,~G.; Li,~Z.; Li,~Z.; Yu,~S.; Zheng,~G. Multiplexed Anticounterfeiting Meta-image Displays with Single-Sized Nanostructures. \emph{Nano Letters} \textbf{2020}, \emph{20}, 1830–1838\relax
\mciteBstWouldAddEndPuncttrue
\mciteSetBstMidEndSepPunct{\mcitedefaultmidpunct}
{\mcitedefaultendpunct}{\mcitedefaultseppunct}\relax
\EndOfBibitem
\bibitem[Singh \latin{et~al.}(2024)Singh, Bhardwaj, Choudhary, Patgiri, Teramoto, and Maji]{Singh2024AppInterfAndMat}
Singh,~S.; Bhardwaj,~S.; Choudhary,~N.; Patgiri,~R.; Teramoto,~Y.; Maji,~P.~K. Stimuli-Responsive Chiral Cellulose Nanocrystals Based Self-Assemblies for Security Measures to Prevent Counterfeiting: A Review. \emph{ACS Applied Materials \& Interfaces} \textbf{2024}, \emph{16}, 41743–41765\relax
\mciteBstWouldAddEndPuncttrue
\mciteSetBstMidEndSepPunct{\mcitedefaultmidpunct}
{\mcitedefaultendpunct}{\mcitedefaultseppunct}\relax
\EndOfBibitem
\bibitem[Khorasaninejad and Capasso(2017)Khorasaninejad, and Capasso]{Khorasaninejad2017S}
Khorasaninejad,~M.; Capasso,~F. {Metalenses: Versatile multifunctional photonic components}. \emph{Science} \textbf{2017}, \emph{358}\relax
\mciteBstWouldAddEndPuncttrue
\mciteSetBstMidEndSepPunct{\mcitedefaultmidpunct}
{\mcitedefaultendpunct}{\mcitedefaultseppunct}\relax
\EndOfBibitem
\bibitem[Arbabi and Faraon(2023)Arbabi, and Faraon]{Arbabi2023NP}
Arbabi,~A.; Faraon,~A. {Advances in optical metalenses}. \emph{Nat. Photonics} \textbf{2023}, \emph{17}, 16--25\relax
\mciteBstWouldAddEndPuncttrue
\mciteSetBstMidEndSepPunct{\mcitedefaultmidpunct}
{\mcitedefaultendpunct}{\mcitedefaultseppunct}\relax
\EndOfBibitem
\bibitem[Zhang \latin{et~al.}(2021)Zhang, Wong, Zeng, Bi, Tai, Dholakia, and Olivo]{Zhang2021Nanophotonics}
Zhang,~S.; Wong,~C.~L.; Zeng,~S.; Bi,~R.; Tai,~K.; Dholakia,~K.; Olivo,~M. {Metasurfaces for biomedical applications: imaging and sensing from a nanophotonics perspective}. \emph{Nanophotonics} \textbf{2021}, \emph{10}, 259--293\relax
\mciteBstWouldAddEndPuncttrue
\mciteSetBstMidEndSepPunct{\mcitedefaultmidpunct}
{\mcitedefaultendpunct}{\mcitedefaultseppunct}\relax
\EndOfBibitem
\bibitem[Tittl \latin{et~al.}(2018)Tittl, Leitis, Liu, Yesilkoy, Choi, Neshev, Kivshar, and Altug]{Tittl2018Science}
Tittl,~A.; Leitis,~A.; Liu,~M.; Yesilkoy,~F.; Choi,~D.-Y.; Neshev,~D.~N.; Kivshar,~Y.~S.; Altug,~H. {Imaging-based molecular barcoding with pixelated dielectric metasurfaces}. \emph{Science} \textbf{2018}, \emph{360}, 1105--1109\relax
\mciteBstWouldAddEndPuncttrue
\mciteSetBstMidEndSepPunct{\mcitedefaultmidpunct}
{\mcitedefaultendpunct}{\mcitedefaultseppunct}\relax
\EndOfBibitem
\bibitem[Krasnok \latin{et~al.}(2018)Krasnok, Tymchenko, and Alù]{Krasnok2018MatToday}
Krasnok,~A.; Tymchenko,~M.; Alù,~A. Nonlinear metasurfaces: a paradigm shift in nonlinear optics. \emph{Materials Today} \textbf{2018}, \emph{21}, 8–21\relax
\mciteBstWouldAddEndPuncttrue
\mciteSetBstMidEndSepPunct{\mcitedefaultmidpunct}
{\mcitedefaultendpunct}{\mcitedefaultseppunct}\relax
\EndOfBibitem
\bibitem[Jangid \latin{et~al.}(2023)Jangid, Richter, Tseng, Sinev, Kruk, Altug, and Kivshar]{Jangid2023AdvMat}
Jangid,~P.; Richter,~F.~U.; Tseng,~M.~L.; Sinev,~I.; Kruk,~S.; Altug,~H.; Kivshar,~Y. Spectral Tuning of High‐Harmonic Generation with Resonance‐Gradient Metasurfaces. \emph{Advanced Materials} \textbf{2023}, \emph{36}\relax
\mciteBstWouldAddEndPuncttrue
\mciteSetBstMidEndSepPunct{\mcitedefaultmidpunct}
{\mcitedefaultendpunct}{\mcitedefaultseppunct}\relax
\EndOfBibitem
\bibitem[Zhou \latin{et~al.}(2024)Zhou, Zhao, He, Huang, Man, and Wan]{Zhou2024Nanophotonics}
Zhou,~H.; Zhao,~C.; He,~C.; Huang,~L.; Man,~T.; Wan,~Y. {Optical computing metasurfaces: applications and advances}. \emph{Nanophotonics} \textbf{2024}, \emph{13}, 419--441\relax
\mciteBstWouldAddEndPuncttrue
\mciteSetBstMidEndSepPunct{\mcitedefaultmidpunct}
{\mcitedefaultendpunct}{\mcitedefaultseppunct}\relax
\EndOfBibitem
\bibitem[Hwang and Davis(2016)Hwang, and Davis]{hwang2016optical}
Hwang,~Y.; Davis,~T.~J. Optical metasurfaces for subwavelength difference operations. \emph{Applied Physics Letters} \textbf{2016}, \emph{109}, 181101\relax
\mciteBstWouldAddEndPuncttrue
\mciteSetBstMidEndSepPunct{\mcitedefaultmidpunct}
{\mcitedefaultendpunct}{\mcitedefaultseppunct}\relax
\EndOfBibitem
\bibitem[Hu \latin{et~al.}(2024)Hu, Mengu, Tzarouchis, Edwards, Engheta, and Ozcan]{Hu2024NatComm}
Hu,~J.; Mengu,~D.; Tzarouchis,~D.~C.; Edwards,~B.; Engheta,~N.; Ozcan,~A. {Diffractive optical computing in free space}. \emph{Nat. Commun.} \textbf{2024}, \emph{15}, 1--21\relax
\mciteBstWouldAddEndPuncttrue
\mciteSetBstMidEndSepPunct{\mcitedefaultmidpunct}
{\mcitedefaultendpunct}{\mcitedefaultseppunct}\relax
\EndOfBibitem
\bibitem[Xie \latin{et~al.}(2021)Xie, Pu, Jin, Xu, Guo, Li, Gao, Ma, and Luo]{Xie2021PRL}
Xie,~X.; Pu,~M.; Jin,~J.; Xu,~M.; Guo,~Y.; Li,~X.; Gao,~P.; Ma,~X.; Luo,~X. {Generalized Pancharatnam-Berry Phase in Rotationally Symmetric Meta-Atoms}. \emph{Phys. Rev. Lett.} \textbf{2021}, \emph{126}, 183902\relax
\mciteBstWouldAddEndPuncttrue
\mciteSetBstMidEndSepPunct{\mcitedefaultmidpunct}
{\mcitedefaultendpunct}{\mcitedefaultseppunct}\relax
\EndOfBibitem
\bibitem[Guo \latin{et~al.}(2022)Guo, Pu, Zhang, Xu, Li, Ma, and Luo]{Guo2022PI}
Guo,~Y.; Pu,~M.; Zhang,~F.; Xu,~M.; Li,~X.; Ma,~X.; Luo,~X. \emph{{Photonics Insights, Vol. 1, Issue 1}}; SPIE, 2022; Vol.~1; p R03\relax
\mciteBstWouldAddEndPuncttrue
\mciteSetBstMidEndSepPunct{\mcitedefaultmidpunct}
{\mcitedefaultendpunct}{\mcitedefaultseppunct}\relax
\EndOfBibitem
\bibitem[Kim \latin{et~al.}(2021)Kim, Rana, Kim, Kim, Badloe, Zubair, Mehmood, and Rho]{Kim2021Jun}
Kim,~J.; Rana,~A.~S.; Kim,~Y.; Kim,~I.; Badloe,~T.; Zubair,~M.; Mehmood,~M.~Q.; Rho,~J. {Chiroptical Metasurfaces: Principles, Classification, and Applications}. \emph{Sensors} \textbf{2021}, \emph{21}, 4381\relax
\mciteBstWouldAddEndPuncttrue
\mciteSetBstMidEndSepPunct{\mcitedefaultmidpunct}
{\mcitedefaultendpunct}{\mcitedefaultseppunct}\relax
\EndOfBibitem
\bibitem[Balthasar~Mueller \latin{et~al.}(2017)Balthasar~Mueller, Rubin, Devlin, Groever, and Capasso]{BalthasarMueller2017Mar}
Balthasar~Mueller,~J.~P.; Rubin,~N.~A.; Devlin,~R.~C.; Groever,~B.; Capasso,~F. {Metasurface Polarization Optics: Independent Phase Control of Arbitrary Orthogonal States of Polarization}. \emph{Phys. Rev. Lett.} \textbf{2017}, \emph{118}, 113901\relax
\mciteBstWouldAddEndPuncttrue
\mciteSetBstMidEndSepPunct{\mcitedefaultmidpunct}
{\mcitedefaultendpunct}{\mcitedefaultseppunct}\relax
\EndOfBibitem
\bibitem[Zhu \latin{et~al.}(2013)Zhu, Cheung, Chung, and Yuk]{Zhu2013IEEE}
Zhu,~H.~L.; Cheung,~S.~W.; Chung,~K.~L.; Yuk,~T.~I. {Linear-to-Circular Polarization Conversion Using Metasurface}. \emph{IEEE Trans. Antennas Propag.} \textbf{2013}, \emph{61}, 4615--4623\relax
\mciteBstWouldAddEndPuncttrue
\mciteSetBstMidEndSepPunct{\mcitedefaultmidpunct}
{\mcitedefaultendpunct}{\mcitedefaultseppunct}\relax
\EndOfBibitem
\bibitem[Teng \latin{et~al.}(2019)Teng, Zhang, Wang, Liu, and Lv]{Teng2019PhotRes}
Teng,~S.; Zhang,~Q.; Wang,~H.; Liu,~L.; Lv,~H. {Conversion between polarization states based on a metasurface}. \emph{Photonics Res.} \textbf{2019}, \emph{7}, 246--250\relax
\mciteBstWouldAddEndPuncttrue
\mciteSetBstMidEndSepPunct{\mcitedefaultmidpunct}
{\mcitedefaultendpunct}{\mcitedefaultseppunct}\relax
\EndOfBibitem
\bibitem[Shah \latin{et~al.}(2022)Shah, Dada, Grant, Cumming, Altuzarra, Nowack, Lyons, Clerici, and Faccio]{Shah2022ACSPhot}
Shah,~Y.~D.; Dada,~A.~C.; Grant,~J.~P.; Cumming,~D. R.~S.; Altuzarra,~C.; Nowack,~T.~S.; Lyons,~A.; Clerici,~M.; Faccio,~D. {An All-Dielectric Metasurface Polarimeter}. \emph{ACS Photonics} \textbf{2022}, \emph{9}, 3245--3252\relax
\mciteBstWouldAddEndPuncttrue
\mciteSetBstMidEndSepPunct{\mcitedefaultmidpunct}
{\mcitedefaultendpunct}{\mcitedefaultseppunct}\relax
\EndOfBibitem
\bibitem[Ding \latin{et~al.}(2018)Ding, Chen, and Bozhevolnyi]{Ding2018ApplSci}
Ding,~F.; Chen,~Y.; Bozhevolnyi,~S.~I. {Metasurface-Based Polarimeters}. \emph{Appl. Sci.} \textbf{2018}, \emph{8}, 594\relax
\mciteBstWouldAddEndPuncttrue
\mciteSetBstMidEndSepPunct{\mcitedefaultmidpunct}
{\mcitedefaultendpunct}{\mcitedefaultseppunct}\relax
\EndOfBibitem
\bibitem[Arbabi \latin{et~al.}(2018)Arbabi, Kamali, Arbabi, and Faraon]{Arbabi2018ACSPhot}
Arbabi,~E.; Kamali,~S.~M.; Arbabi,~A.; Faraon,~A. {Full-Stokes Imaging Polarimetry Using Dielectric Metasurfaces}. \emph{ACS Photonics} \textbf{2018}, \emph{5}, 3132--3140\relax
\mciteBstWouldAddEndPuncttrue
\mciteSetBstMidEndSepPunct{\mcitedefaultmidpunct}
{\mcitedefaultendpunct}{\mcitedefaultseppunct}\relax
\EndOfBibitem
\bibitem[Tang and Cohen(2011)Tang, and Cohen]{Tang2011Apr}
Tang,~Y.; Cohen,~A.~E. {Enhanced Enantioselectivity in Excitation of Chiral Molecules by Superchiral Light}. \emph{Science} \textbf{2011}, \emph{332}, 333--336\relax
\mciteBstWouldAddEndPuncttrue
\mciteSetBstMidEndSepPunct{\mcitedefaultmidpunct}
{\mcitedefaultendpunct}{\mcitedefaultseppunct}\relax
\EndOfBibitem
\bibitem[Mohammadi \latin{et~al.}(2018)Mohammadi, Tsakmakidis, Askarpour, Dehkhoda, Tavakoli, and Altug]{Mohammadi2018Jul}
Mohammadi,~E.; Tsakmakidis,~K.~L.; Askarpour,~A.~N.; Dehkhoda,~P.; Tavakoli,~A.; Altug,~H. {Nanophotonic Platforms for Enhanced Chiral Sensing}. \emph{ACS Photonics} \textbf{2018}, \emph{5}, 2669--2675\relax
\mciteBstWouldAddEndPuncttrue
\mciteSetBstMidEndSepPunct{\mcitedefaultmidpunct}
{\mcitedefaultendpunct}{\mcitedefaultseppunct}\relax
\EndOfBibitem
\bibitem[Mohammadi \latin{et~al.}(2019)Mohammadi, Tavakoli, Dehkhoda, Jahani, Tsakmakidis, Tittl, and Altug]{Mohammadi2019Aug}
Mohammadi,~E.; Tavakoli,~A.; Dehkhoda,~P.; Jahani,~Y.; Tsakmakidis,~K.~L.; Tittl,~A.; Altug,~H. {Accessible Superchiral Near-Fields Driven by Tailored Electric and Magnetic Resonances in All-Dielectric Nanostructures}. \emph{ACS Photonics} \textbf{2019}, \emph{6}, 1939--1946\relax
\mciteBstWouldAddEndPuncttrue
\mciteSetBstMidEndSepPunct{\mcitedefaultmidpunct}
{\mcitedefaultendpunct}{\mcitedefaultseppunct}\relax
\EndOfBibitem
\bibitem[Garcia-Santiago \latin{et~al.}(2022)Garcia-Santiago, Hammerschmidt, Sachs, Burger, Kwon, Kn{\ifmmode\ddot{o}\else\"{o}\fi}ller, Arens, Fischer, Fernandez-Corbaton, and Rockstuhl]{Garcia-Santiago2022Jun}
Garcia-Santiago,~X.; Hammerschmidt,~M.; Sachs,~J.; Burger,~S.; Kwon,~H.; Kn{\ifmmode\ddot{o}\else\"{o}\fi}ller,~M.; Arens,~T.; Fischer,~P.; Fernandez-Corbaton,~I.; Rockstuhl,~C. {Toward Maximally Electromagnetically Chiral Scatterers at Optical Frequencies}. \emph{ACS Photonics} \textbf{2022}, \emph{9}, 1954--1964\relax
\mciteBstWouldAddEndPuncttrue
\mciteSetBstMidEndSepPunct{\mcitedefaultmidpunct}
{\mcitedefaultendpunct}{\mcitedefaultseppunct}\relax
\EndOfBibitem
\bibitem[Khaliq \latin{et~al.}(2023)Khaliq, Nauman, Lee, and Kim]{Khaliq2023AdvOptMat}
Khaliq,~H.~S.; Nauman,~A.; Lee,~J.; Kim,~H. Recent Progress on Plasmonic and Dielectric Chiral Metasurfaces: Fundamentals, Design Strategies, and Implementation. \emph{Advanced Optical Materials} \textbf{2023}, \emph{11}\relax
\mciteBstWouldAddEndPuncttrue
\mciteSetBstMidEndSepPunct{\mcitedefaultmidpunct}
{\mcitedefaultendpunct}{\mcitedefaultseppunct}\relax
\EndOfBibitem
\bibitem[Deng \latin{et~al.}(2024)Deng, Li, Hu, Li, Li, and Deng]{Deng2024NatNanoPhot}
Deng,~Q.-M.; Li,~X.; Hu,~M.-X.; Li,~F.-J.; Li,~X.; Deng,~Z.-L. {Advances on broadband and resonant chiral metasurfaces}. \emph{npj Nanophoton.} \textbf{2024}, \emph{1}, 1--22\relax
\mciteBstWouldAddEndPuncttrue
\mciteSetBstMidEndSepPunct{\mcitedefaultmidpunct}
{\mcitedefaultendpunct}{\mcitedefaultseppunct}\relax
\EndOfBibitem
\bibitem[Plum \latin{et~al.}(2009)Plum, Fedotov, and Zheludev]{Plum2009APL}
Plum,~E.; Fedotov,~V.~A.; Zheludev,~N.~I. {Planar metamaterial with transmission and reflection that depend on the direction of incidence}. \emph{Appl. Phys. Lett.} \textbf{2009}, \emph{94}, 131901\relax
\mciteBstWouldAddEndPuncttrue
\mciteSetBstMidEndSepPunct{\mcitedefaultmidpunct}
{\mcitedefaultendpunct}{\mcitedefaultseppunct}\relax
\EndOfBibitem
\bibitem[Koshelev \latin{et~al.}(2024)Koshelev, Toftul, Hwang, and Kivshar]{Koshelev2024JO}
Koshelev,~K.; Toftul,~I.; Hwang,~Y.; Kivshar,~Y. Scattering Matrix for Chiral Harmonic Generation and Frequency Mixing in Nonlinear Metasurfaces. \emph{J. Opt.} \textbf{2024}, \emph{26}, 055003\relax
\mciteBstWouldAddEndPuncttrue
\mciteSetBstMidEndSepPunct{\mcitedefaultmidpunct}
{\mcitedefaultendpunct}{\mcitedefaultseppunct}\relax
\EndOfBibitem
\bibitem[Arteaga \latin{et~al.}(2016)Arteaga, Sancho-Parramon, Nichols, Maoz, Canillas, Bosch, Markovich, and Kahr]{Arteaga2016Feb}
Arteaga,~O.; Sancho-Parramon,~J.; Nichols,~S.; Maoz,~B.~M.; Canillas,~A.; Bosch,~S.; Markovich,~G.; Kahr,~B. {Relation between 2D/3D chirality and the appearance of chiroptical effects in real nanostructures}. \emph{Opt. Express} \textbf{2016}, \emph{24}, 2242--2252\relax
\mciteBstWouldAddEndPuncttrue
\mciteSetBstMidEndSepPunct{\mcitedefaultmidpunct}
{\mcitedefaultendpunct}{\mcitedefaultseppunct}\relax
\EndOfBibitem
\bibitem[Goerlitzer \latin{et~al.}(2020)Goerlitzer, Mohammadi, Nechayev, Volk, Rey, Banzer, Karg, and Vogel]{Goerlitzer2020Jun}
Goerlitzer,~E. S.~A.; Mohammadi,~R.; Nechayev,~S.; Volk,~K.; Rey,~M.; Banzer,~P.; Karg,~M.; Vogel,~N. {Chiral Surface Lattice Resonances}. \emph{Adv. Mater.} \textbf{2020}, \emph{32}, 2001330\relax
\mciteBstWouldAddEndPuncttrue
\mciteSetBstMidEndSepPunct{\mcitedefaultmidpunct}
{\mcitedefaultendpunct}{\mcitedefaultseppunct}\relax
\EndOfBibitem
\bibitem[Gorkunov \latin{et~al.}(2020)Gorkunov, Antonov, and Kivshar]{Gorkunov2020PRL}
Gorkunov,~M.~V.; Antonov,~A.~A.; Kivshar,~Y.~S. Metasurfaces with {{Maximum Chirality Empowered}} by {{Bound States}} in the {{Continuum}}. \emph{Phys. Rev. Lett.} \textbf{2020}, \emph{125}, 093903\relax
\mciteBstWouldAddEndPuncttrue
\mciteSetBstMidEndSepPunct{\mcitedefaultmidpunct}
{\mcitedefaultendpunct}{\mcitedefaultseppunct}\relax
\EndOfBibitem
\bibitem[Tanaka \latin{et~al.}(2020)Tanaka, Arslan, Fasold, Steinert, Sautter, Falkner, Pertsch, Decker, and Staude]{Tanaka2020ACSNano}
Tanaka,~K.; Arslan,~D.; Fasold,~S.; Steinert,~M.; Sautter,~J.; Falkner,~M.; Pertsch,~T.; Decker,~M.; Staude,~I. {Chiral Bilayer All-Dielectric Metasurfaces}. \emph{ACS Nano} \textbf{2020}, \emph{14}, 15926--15935\relax
\mciteBstWouldAddEndPuncttrue
\mciteSetBstMidEndSepPunct{\mcitedefaultmidpunct}
{\mcitedefaultendpunct}{\mcitedefaultseppunct}\relax
\EndOfBibitem
\bibitem[Kosters \latin{et~al.}(2017)Kosters, de~Hoogh, Zeijlemaker, Acar, Rotenberg, and Kuipers]{Kosters2017ACSPhot}
Kosters,~D.; de~Hoogh,~A.; Zeijlemaker,~H.; Acar,~H.; Rotenberg,~N.; Kuipers,~L. {Core{\textendash}Shell Plasmonic Nanohelices}. \emph{ACS Photonics} \textbf{2017}, \emph{4}, 1858--1863\relax
\mciteBstWouldAddEndPuncttrue
\mciteSetBstMidEndSepPunct{\mcitedefaultmidpunct}
{\mcitedefaultendpunct}{\mcitedefaultseppunct}\relax
\EndOfBibitem
\bibitem[Esposito \latin{et~al.}(2015)Esposito, Tasco, Cuscun{\ifmmode\grave{a}\else\`{a}\fi}, Todisco, Benedetti, Tarantini, Giorgi, Sanvitto, and Passaseo]{Esposito2015ACSPhot}
Esposito,~M.; Tasco,~V.; Cuscun{\ifmmode\grave{a}\else\`{a}\fi},~M.; Todisco,~F.; Benedetti,~A.; Tarantini,~I.; Giorgi,~M.~D.; Sanvitto,~D.; Passaseo,~A. {Nanoscale 3D Chiral Plasmonic Helices with Circular Dichroism at Visible Frequencies}. \emph{ACS Photonics} \textbf{2015}, \emph{2}, 105--114\relax
\mciteBstWouldAddEndPuncttrue
\mciteSetBstMidEndSepPunct{\mcitedefaultmidpunct}
{\mcitedefaultendpunct}{\mcitedefaultseppunct}\relax
\EndOfBibitem
\bibitem[Kaschke \latin{et~al.}(2015)Kaschke, Blume, Wu, Thiel, Bade, Yang, and Wegener]{Kaschke2015AOM}
Kaschke,~J.; Blume,~L.; Wu,~L.; Thiel,~M.; Bade,~K.; Yang,~Z.; Wegener,~M. {A Helical Metamaterial for Broadband Circular Polarization Conversion}. \emph{Adv. Opt. Mater.} \textbf{2015}, \emph{3}, 1411--1417\relax
\mciteBstWouldAddEndPuncttrue
\mciteSetBstMidEndSepPunct{\mcitedefaultmidpunct}
{\mcitedefaultendpunct}{\mcitedefaultseppunct}\relax
\EndOfBibitem
\bibitem[Shi \latin{et~al.}(2022)Shi, Deng, Geng, Zeng, Zeng, Hu, Overvig, Li, Qiu, Al{\ifmmode\grave{u}\else\`{u}\fi}, Kivshar, and Li]{Shi2022NatCom}
Shi,~T.; Deng,~Z.-L.; Geng,~G.; Zeng,~X.; Zeng,~Y.; Hu,~G.; Overvig,~A.; Li,~J.; Qiu,~C.-W.; Al{\ifmmode\grave{u}\else\`{u}\fi},~A.; Kivshar,~Y.~S.; Li,~X. {Planar chiral metasurfaces with maximal and tunable chiroptical response driven by bound states in the continuum}. \emph{Nat. Commun.} \textbf{2022}, \emph{13}, 1--8\relax
\mciteBstWouldAddEndPuncttrue
\mciteSetBstMidEndSepPunct{\mcitedefaultmidpunct}
{\mcitedefaultendpunct}{\mcitedefaultseppunct}\relax
\EndOfBibitem
\bibitem[Wang \latin{et~al.}(2023)Wang, Wang, Sun, Hu, and Wang]{Wang2023NanoPhot}
Wang,~R.; Wang,~C.; Sun,~T.; Hu,~X.; Wang,~C. Simultaneous broadband and high circular dichroism with two-dimensional all-dielectric chiral metasurface. \emph{Nanophotonics} \textbf{2023}, \emph{12}, 4043–4053\relax
\mciteBstWouldAddEndPuncttrue
\mciteSetBstMidEndSepPunct{\mcitedefaultmidpunct}
{\mcitedefaultendpunct}{\mcitedefaultseppunct}\relax
\EndOfBibitem
\bibitem[Koshelev \latin{et~al.}(2023)Koshelev, Tang, Hu, Kravchenko, Li, and Kivshar]{Koshelev2023ACSPhot}
Koshelev,~K.; Tang,~Y.; Hu,~Z.; Kravchenko,~I.~I.; Li,~G.; Kivshar,~Y. {Resonant Chiral Effects in Nonlinear Dielectric Metasurfaces}. \emph{ACS Photonics} \textbf{2023}, \emph{10}, 298--306\relax
\mciteBstWouldAddEndPuncttrue
\mciteSetBstMidEndSepPunct{\mcitedefaultmidpunct}
{\mcitedefaultendpunct}{\mcitedefaultseppunct}\relax
\EndOfBibitem
\bibitem[Tonkaev \latin{et~al.}(2024)Tonkaev, Toftul, Lu, Qin, Qiu, Yang, Koshelev, Lu, and Kivshar]{Tonkaev2024NL}
Tonkaev,~P.; Toftul,~I.; Lu,~Z.; Qin,~H.; Qiu,~S.; Yang,~W.; Koshelev,~K.; Lu,~Y.; Kivshar,~Y. {Nonlinear Chiral Metasurfaces Based on Structured van der Waals Materials}. \emph{Nano Lett.} \textbf{2024}, \emph{24}, 10577--10582\relax
\mciteBstWouldAddEndPuncttrue
\mciteSetBstMidEndSepPunct{\mcitedefaultmidpunct}
{\mcitedefaultendpunct}{\mcitedefaultseppunct}\relax
\EndOfBibitem
\bibitem[Toftul \latin{et~al.}(2024)Toftul, Tonkaev, Koshelev, Lai, Song, Gorkunov, and Kivshar]{Toftul2024PRL}
Toftul,~I.; Tonkaev,~P.; Koshelev,~K.; Lai,~F.; Song,~Q.; Gorkunov,~M.; Kivshar,~Y. {Chiral Dichroism in Resonant Metasurfaces with Monoclinic Lattices}. \emph{Phys. Rev. Lett.} \textbf{2024}, \emph{133}, 216901\relax
\mciteBstWouldAddEndPuncttrue
\mciteSetBstMidEndSepPunct{\mcitedefaultmidpunct}
{\mcitedefaultendpunct}{\mcitedefaultseppunct}\relax
\EndOfBibitem
\bibitem[Lyu \latin{et~al.}(2023)Lyu, Li, Jia, Li, Yang, Cao, Kou, Liu, Cao, Li, and Shi]{Lyu2023LPR}
Lyu,~B.; Li,~Y.; Jia,~Q.; Li,~H.; Yang,~G.; Cao,~F.; Kou,~S.; Liu,~D.; Cao,~T.; Li,~G.; Shi,~J. {Manipulating the Chirality of Moir{\ifmmode\acute{e}\else\'{e}\fi} Metasurface by Symmetry Breaking}. \emph{Laser Photonics Rev.} \textbf{2023}, \emph{17}, 2201004\relax
\mciteBstWouldAddEndPuncttrue
\mciteSetBstMidEndSepPunct{\mcitedefaultmidpunct}
{\mcitedefaultendpunct}{\mcitedefaultseppunct}\relax
\EndOfBibitem
\bibitem[Wu \latin{et~al.}(2018)Wu, Liu, Hill, and Zheng]{Wu2018Nanoscale}
Wu,~Z.; Liu,~Y.; Hill,~E.~H.; Zheng,~Y. {Chiral metamaterials via Moir{\ifmmode\acute{e}\else\'{e}\fi} stacking}. \emph{Nanoscale} \textbf{2018}, \emph{10}, 18096--18112\relax
\mciteBstWouldAddEndPuncttrue
\mciteSetBstMidEndSepPunct{\mcitedefaultmidpunct}
{\mcitedefaultendpunct}{\mcitedefaultseppunct}\relax
\EndOfBibitem
\bibitem[Han \latin{et~al.}(2023)Han, Wang, Sun, Wang, and Tang]{Han2023AM}
Han,~Z.; Wang,~F.; Sun,~J.; Wang,~X.; Tang,~Z. {Recent Advances in Ultrathin Chiral Metasurfaces by Twisted Stacking}. \emph{Adv. Mater.} \textbf{2023}, \emph{35}, 2206141\relax
\mciteBstWouldAddEndPuncttrue
\mciteSetBstMidEndSepPunct{\mcitedefaultmidpunct}
{\mcitedefaultendpunct}{\mcitedefaultseppunct}\relax
\EndOfBibitem
\bibitem[Asadchy \latin{et~al.}(2018)Asadchy, D{\ifmmode\acute{\imath}\else\'{\i}\fi}az-Rubio, and Tretyakov]{Asadchy2018Jun}
Asadchy,~V.~S.; D{\ifmmode\acute{\imath}\else\'{\i}\fi}az-Rubio,~A.; Tretyakov,~S.~A. {Bianisotropic metasurfaces: physics and applications}. \emph{Nanophotonics} \textbf{2018}, \emph{7}, 1069--1094\relax
\mciteBstWouldAddEndPuncttrue
\mciteSetBstMidEndSepPunct{\mcitedefaultmidpunct}
{\mcitedefaultendpunct}{\mcitedefaultseppunct}\relax
\EndOfBibitem
\bibitem[Cao \latin{et~al.}(2016)Cao, wei Wei, and Li]{Cao2016OptMatExp}
Cao,~T.; wei Wei,~C.; Li,~Y. Dual-band strong extrinsic 2D chirality in a highly symmetric metal-dielectric-metal achiral metasurface. \emph{Opt. Mater. Express} \textbf{2016}, \emph{6}, 303--311\relax
\mciteBstWouldAddEndPuncttrue
\mciteSetBstMidEndSepPunct{\mcitedefaultmidpunct}
{\mcitedefaultendpunct}{\mcitedefaultseppunct}\relax
\EndOfBibitem
\bibitem[Hwang \latin{et~al.}(2017)Hwang, Hopkins, Wang, Mitchell, Davis, Lin, and Yuan]{hwang2017optical}
Hwang,~Y.; Hopkins,~B.; Wang,~D.; Mitchell,~A.; Davis,~T.~J.; Lin,~J.; Yuan,~X.-C. Optical Chirality from Dark-Field Illumination of Planar Plasmonic Nanostructures. \emph{Laser \& Photonics Reviews} \textbf{2017}, \emph{11}, 1700216\relax
\mciteBstWouldAddEndPuncttrue
\mciteSetBstMidEndSepPunct{\mcitedefaultmidpunct}
{\mcitedefaultendpunct}{\mcitedefaultseppunct}\relax
\EndOfBibitem
\bibitem[Nicolas \latin{et~al.}(2023)Nicolas, Walmsness, Amboli, Zhang, Demesy, Bonod, Boujday, Kildemo, and Gallas]{Nicolas2023ApplOptMat}
Nicolas,~M.; Walmsness,~P.~M.; Amboli,~J.; Zhang,~L.; Demesy,~G.; Bonod,~N.; Boujday,~S.; Kildemo,~M.; Gallas,~B. True Circular Dichroism in Optically Active Achiral Metasurfaces and Its Relation to Chiral Near-Fields. \emph{ACS Applied Optical Materials} \textbf{2023}, \emph{1}, 1360–1366\relax
\mciteBstWouldAddEndPuncttrue
\mciteSetBstMidEndSepPunct{\mcitedefaultmidpunct}
{\mcitedefaultendpunct}{\mcitedefaultseppunct}\relax
\EndOfBibitem
\bibitem[Graf \latin{et~al.}(2019)Graf, Feis, Garcia-Santiago, Wegener, Rockstuhl, and Fernandez-Corbaton]{Graf2019Feb}
Graf,~F.; Feis,~J.; Garcia-Santiago,~X.; Wegener,~M.; Rockstuhl,~C.; Fernandez-Corbaton,~I. {Achiral, Helicity Preserving, and Resonant Structures for Enhanced Sensing of Chiral Molecules}. \emph{ACS Photonics} \textbf{2019}, \emph{6}, 482--491\relax
\mciteBstWouldAddEndPuncttrue
\mciteSetBstMidEndSepPunct{\mcitedefaultmidpunct}
{\mcitedefaultendpunct}{\mcitedefaultseppunct}\relax
\EndOfBibitem
\bibitem[Gilroy \latin{et~al.}(2019)Gilroy, Hashiyada, Endo, Karimullah, Barron, Okamoto, Togawa, and Kadodwala]{Gilroy2019Jun}
Gilroy,~C.; Hashiyada,~S.; Endo,~K.; Karimullah,~A.~S.; Barron,~L.~D.; Okamoto,~H.; Togawa,~Y.; Kadodwala,~M. {Roles of Superchirality and Interference in Chiral Plasmonic Biodetection}. \emph{J. Phys. Chem. C} \textbf{2019}, \emph{123}, 15195--15203\relax
\mciteBstWouldAddEndPuncttrue
\mciteSetBstMidEndSepPunct{\mcitedefaultmidpunct}
{\mcitedefaultendpunct}{\mcitedefaultseppunct}\relax
\EndOfBibitem
\bibitem[Both \latin{et~al.}(2022)Both, Sch{\ifmmode\ddot{a}\else\"{a}\fi}ferling, Sterl, Muljarov, Giessen, and Weiss]{Both2022Feb}
Both,~S.; Sch{\ifmmode\ddot{a}\else\"{a}\fi}ferling,~M.; Sterl,~F.; Muljarov,~E.~A.; Giessen,~H.; Weiss,~T. {Nanophotonic Chiral Sensing: How Does It Actually Work?} \emph{ACS Nano} \textbf{2022}, \emph{16}, 2822--2832\relax
\mciteBstWouldAddEndPuncttrue
\mciteSetBstMidEndSepPunct{\mcitedefaultmidpunct}
{\mcitedefaultendpunct}{\mcitedefaultseppunct}\relax
\EndOfBibitem
\bibitem[Volkov \latin{et~al.}(2009)Volkov, Dolgaleva, Boyd, Jefimovs, Turunen, Svirko, Canfield, and Kauranen]{Volkov2009Apr}
Volkov,~S.~N.; Dolgaleva,~K.; Boyd,~R.~W.; Jefimovs,~K.; Turunen,~J.; Svirko,~Y.; Canfield,~B.~K.; Kauranen,~M. {Optical activity in diffraction from a planar array of achiral nanoparticles}. \emph{Phys. Rev. A} \textbf{2009}, \emph{79}, 043819\relax
\mciteBstWouldAddEndPuncttrue
\mciteSetBstMidEndSepPunct{\mcitedefaultmidpunct}
{\mcitedefaultendpunct}{\mcitedefaultseppunct}\relax
\EndOfBibitem
\bibitem[Meng \latin{et~al.}(2022)Meng, Zhang, Liu, Li, Sun, Lai, and Yu]{Meng2022Oct}
Meng,~J.; Zhang,~Z.; Liu,~W.; Li,~Y.; Sun,~Y.; Lai,~Z.; Yu,~T. {Angle-selective chiral absorption induced by diffractive coupling in metasurfaces}. \emph{Opt. Lett.} \textbf{2022}, \emph{47}, 5385--5388\relax
\mciteBstWouldAddEndPuncttrue
\mciteSetBstMidEndSepPunct{\mcitedefaultmidpunct}
{\mcitedefaultendpunct}{\mcitedefaultseppunct}\relax
\EndOfBibitem
\bibitem[Movsesyan \latin{et~al.}(2022)Movsesyan, Besteiro, Kong, Wang, and Govorov]{Movsesyan2022AOM}
Movsesyan,~A.; Besteiro,~L.~V.; Kong,~X.-T.; Wang,~Z.; Govorov,~A.~O. {Engineering Strongly Chiral Plasmonic Lattices with Achiral Unit Cells for Sensing and Photodetection}. \emph{Adv. Opt. Mater.} \textbf{2022}, \emph{10}, 2101943\relax
\mciteBstWouldAddEndPuncttrue
\mciteSetBstMidEndSepPunct{\mcitedefaultmidpunct}
{\mcitedefaultendpunct}{\mcitedefaultseppunct}\relax
\EndOfBibitem
\bibitem[{\ifmmode\acute{A}\else\'{A}\fi}valos Ovando \latin{et~al.}(2022){\ifmmode\acute{A}\else\'{A}\fi}valos Ovando, Santiago, Movsesyan, Kong, Yu, Besteiro, Khorashad, Okamoto, Slocik, Correa-Duarte, Comesa{\ifmmode\tilde{n}\else\~{n}\fi}a-Hermo, Liedl, Wang, Markovich, Burger, and Govorov]{Avalos-Ovando2022ACSPhot}
{\ifmmode\acute{A}\else\'{A}\fi}valos Ovando,~O.; Santiago,~E.~Y.; Movsesyan,~A.; Kong,~X.-T.; Yu,~P.; Besteiro,~L.~V.; Khorashad,~L.~K.; Okamoto,~H.; Slocik,~J.~M.; Correa-Duarte,~M.~A.; Comesa{\ifmmode\tilde{n}\else\~{n}\fi}a-Hermo,~M.; Liedl,~T.; Wang,~Z.; Markovich,~G.; Burger,~S.; Govorov,~A.~O. {Chiral Bioinspired Plasmonics: A Paradigm Shift for Optical Activity and Photochemistry}. \emph{ACS Photonics} \textbf{2022}, \emph{9}, 2219--2236\relax
\mciteBstWouldAddEndPuncttrue
\mciteSetBstMidEndSepPunct{\mcitedefaultmidpunct}
{\mcitedefaultendpunct}{\mcitedefaultseppunct}\relax
\EndOfBibitem
\bibitem[Gryb \latin{et~al.}(2023)Gryb, Wendisch, Aigner, G{\ifmmode\ddot{o}\else\"{o}\fi}lz, Tittl, de~S.~Menezes, and Maier]{Gryb2023NL}
Gryb,~D.; Wendisch,~F.~J.; Aigner,~A.; G{\ifmmode\ddot{o}\else\"{o}\fi}lz,~T.; Tittl,~A.; de~S.~Menezes,~L.; Maier,~S.~A. {Two-Dimensional Chiral Metasurfaces Obtained by Geometrically Simple Meta-atom Rotations}. \emph{Nano Lett.} \textbf{2023}, \emph{23}, 8891--8897\relax
\mciteBstWouldAddEndPuncttrue
\mciteSetBstMidEndSepPunct{\mcitedefaultmidpunct}
{\mcitedefaultendpunct}{\mcitedefaultseppunct}\relax
\EndOfBibitem
\bibitem[Gorkunov \latin{et~al.}(2024)Gorkunov, Antonov, Mamonova, Muljarov, and Kivshar]{Gorkunov2024AdvOptMat}
Gorkunov,~M.~V.; Antonov,~A.~A.; Mamonova,~A.~V.; Muljarov,~E.~A.; Kivshar,~Y. {Substrate-Induced Maximum Optical Chirality of Planar Dielectric Structures}. \emph{Adv. Opt. Mater.} \textbf{2024}, \emph{n/a}, 2402133\relax
\mciteBstWouldAddEndPuncttrue
\mciteSetBstMidEndSepPunct{\mcitedefaultmidpunct}
{\mcitedefaultendpunct}{\mcitedefaultseppunct}\relax
\EndOfBibitem
\bibitem[Nechayev \latin{et~al.}(2019)Nechayev, Barczyk, Mick, and Banzer]{Nechayev2019ACSPhot}
Nechayev,~S.; Barczyk,~R.; Mick,~U.; Banzer,~P. {Substrate-Induced Chirality in an Individual Nanostructure}. \emph{ACS Photonics} \textbf{2019}, \emph{6}, 1876--1881\relax
\mciteBstWouldAddEndPuncttrue
\mciteSetBstMidEndSepPunct{\mcitedefaultmidpunct}
{\mcitedefaultendpunct}{\mcitedefaultseppunct}\relax
\EndOfBibitem
\bibitem[Kittel(2005)]{Kittel2005}
Kittel,~C. \emph{Introduction to Solid State Physics}, 8th ed.; Wiley: Hoboken, NJ, 2005\relax
\mciteBstWouldAddEndPuncttrue
\mciteSetBstMidEndSepPunct{\mcitedefaultmidpunct}
{\mcitedefaultendpunct}{\mcitedefaultseppunct}\relax
\EndOfBibitem
\bibitem[Wakabayashi \latin{et~al.}(2014)Wakabayashi, Yokojima, Fukaminato, Shiino, Irie, and Nakamura]{Wakabayashi2014JPCA}
Wakabayashi,~M.; Yokojima,~S.; Fukaminato,~T.; Shiino,~K.; Irie,~M.; Nakamura,~S. {Anisotropic Dissymmetry Factor, g: Theoretical Investigation on Single Molecule Chiroptical Spectroscopy}. \emph{J. Phys. Chem. A} \textbf{2014}, \emph{118}, 5046--5057\relax
\mciteBstWouldAddEndPuncttrue
\mciteSetBstMidEndSepPunct{\mcitedefaultmidpunct}
{\mcitedefaultendpunct}{\mcitedefaultseppunct}\relax
\EndOfBibitem
\bibitem[Berova \latin{et~al.}(2007)Berova, Bari, and Pescitelli]{Berova2007CSR}
Berova,~N.; Bari,~L.~D.; Pescitelli,~G. {Application of electronic circular dichroism in configurational and conformational analysis of organic compounds}. \emph{Chem. Soc. Rev.} \textbf{2007}, \emph{36}, 914--931\relax
\mciteBstWouldAddEndPuncttrue
\mciteSetBstMidEndSepPunct{\mcitedefaultmidpunct}
{\mcitedefaultendpunct}{\mcitedefaultseppunct}\relax
\EndOfBibitem
\bibitem[Shalin \latin{et~al.}(2023)Shalin, Valero, and Miroshnichenko]{Shalin2023}
Shalin,~A.~S.; Valero,~A.~C.; Miroshnichenko,~A. \emph{All-Dielectric Nanophotonics}; Elsevier: San Diego, 2023\relax
\mciteBstWouldAddEndPuncttrue
\mciteSetBstMidEndSepPunct{\mcitedefaultmidpunct}
{\mcitedefaultendpunct}{\mcitedefaultseppunct}\relax
\EndOfBibitem
\bibitem[Kondratov \latin{et~al.}(2016)Kondratov, Gorkunov, Darinskii, Gainutdinov, Rogov, Ezhov, and Artemov]{Kondratov2016PRB}
Kondratov,~A.~V.; Gorkunov,~M.~V.; Darinskii,~A.~N.; Gainutdinov,~R.~V.; Rogov,~O.~Y.; Ezhov,~A.~A.; Artemov,~V.~V. {Extreme optical chirality of plasmonic nanohole arrays due to chiral Fano resonance}. \emph{Phys. Rev. B} \textbf{2016}, \emph{93}, 195418\relax
\mciteBstWouldAddEndPuncttrue
\mciteSetBstMidEndSepPunct{\mcitedefaultmidpunct}
{\mcitedefaultendpunct}{\mcitedefaultseppunct}\relax
\EndOfBibitem
\bibitem[Leitis \latin{et~al.}(2019)Leitis, Tittl, Liu, Lee, Gu, Kivshar, and Altug]{Leitis2019SciAdv}
Leitis,~A.; Tittl,~A.; Liu,~M.; Lee,~B.~H.; Gu,~M.~B.; Kivshar,~Y.~S.; Altug,~H. {Angle-multiplexed all-dielectric metasurfaces for broadband molecular fingerprint retrieval}. \emph{Sci. Adv.} \textbf{2019}, \emph{5}, eaaw2871\relax
\mciteBstWouldAddEndPuncttrue
\mciteSetBstMidEndSepPunct{\mcitedefaultmidpunct}
{\mcitedefaultendpunct}{\mcitedefaultseppunct}\relax
\EndOfBibitem
\bibitem[Richter \latin{et~al.}(2024)Richter, Sinev, Zhou, Leitis, Oh, Tseng, Kivshar, and Altug]{Richter2024Jun}
Richter,~F.~U.; Sinev,~I.; Zhou,~S.; Leitis,~A.; Oh,~S.-H.; Tseng,~M.~L.; Kivshar,~Y.; Altug,~H. {Gradient High-Q Dielectric Metasurfaces for Broadband Sensing and Control of Vibrational Light-Matter Coupling}. \emph{Adv. Mater.} \textbf{2024}, \emph{36}, 2314279\relax
\mciteBstWouldAddEndPuncttrue
\mciteSetBstMidEndSepPunct{\mcitedefaultmidpunct}
{\mcitedefaultendpunct}{\mcitedefaultseppunct}\relax
\EndOfBibitem
\bibitem[Gromyko \latin{et~al.}(2024)Gromyko, An, Gorelik, Xu, Lim, Lee, Tjiptoharsono, Tan, Qiu, Dong, and Wu]{Gromyko2024NC}
Gromyko,~D.; An,~S.; Gorelik,~S.; Xu,~J.; Lim,~L.~J.; Lee,~H. Y.~L.; Tjiptoharsono,~F.; Tan,~Z.-K.; Qiu,~C.-W.; Dong,~Z.; Wu,~L. {Unidirectional Chiral Emission via Twisted Bi-layer Metasurfaces}. \emph{Nat. Commun.} \textbf{2024}, \emph{15}, 1--10\relax
\mciteBstWouldAddEndPuncttrue
\mciteSetBstMidEndSepPunct{\mcitedefaultmidpunct}
{\mcitedefaultendpunct}{\mcitedefaultseppunct}\relax
\EndOfBibitem
\bibitem[Gromyko \latin{et~al.}(2024)Gromyko, Loh, Feng, Qiu, and Wu]{Gromyko2024arXiv}
Gromyko,~D.; Loh,~J.~S.; Feng,~J.; Qiu,~C.-W.; Wu,~L. {Enabling all-to-circular polarization upconversion by nonlinear chiral metasurfaces with rotational symmetry}. \emph{arXiv} \textbf{2024}, \relax
\mciteBstWouldAddEndPunctfalse
\mciteSetBstMidEndSepPunct{\mcitedefaultmidpunct}
{}{\mcitedefaultseppunct}\relax
\EndOfBibitem
\bibitem[Fan \latin{et~al.}(2020)Fan, Liu, Zhang, Zhu, Wang, Lin, Yan, Chen, Lezec, Lu, Agrawal, and Xu]{Fan2020Dec}
Fan,~Q.; Liu,~M.; Zhang,~C.; Zhu,~W.; Wang,~Y.; Lin,~P.; Yan,~F.; Chen,~L.; Lezec,~H.~J.; Lu,~Y.; Agrawal,~A.; Xu,~T. {Independent Amplitude Control of Arbitrary Orthogonal States of Polarization via Dielectric Metasurfaces}. \emph{Phys. Rev. Lett.} \textbf{2020}, \emph{125}, 267402\relax
\mciteBstWouldAddEndPuncttrue
\mciteSetBstMidEndSepPunct{\mcitedefaultmidpunct}
{\mcitedefaultendpunct}{\mcitedefaultseppunct}\relax
\EndOfBibitem
\bibitem[Wu \latin{et~al.}(2022)Wu, Wang, Sun, Wu, Shi, and Wu]{Wu2022AdvComposHybMater}
Wu,~B.; Wang,~M.; Sun,~Y.; Wu,~F.; Shi,~Z.; Wu,~X. Near-infrared chirality of plasmonic metasurfaces with gold rectangular holes. \emph{Advanced Composites and Hybrid Materials} \textbf{2022}, \emph{5}, 2527–2535\relax
\mciteBstWouldAddEndPuncttrue
\mciteSetBstMidEndSepPunct{\mcitedefaultmidpunct}
{\mcitedefaultendpunct}{\mcitedefaultseppunct}\relax
\EndOfBibitem
\end{mcitethebibliography}
\end{document}